\documentclass[aps,showpacs,pra,twocolumn,superscriptaddress]{revtex4}
\usepackage{graphicx}
\usepackage{amssymb}
\usepackage{epstopdf}
\usepackage{color, amsmath, bm}
\usepackage[hidelinks]{hyperref}
\usepackage{bbold}
\newcommand{\be}{\begin{equation}}
\newcommand{\ee}{  \end{equation}}
\newcommand{\ba}{\begin{eqnarray}}
\newcommand{\ea}{  \end{eqnarray}}

\begin{document}

\title{Enhancement of   quantum correlations  and geometric phase for a driven bipartite  quantum system in a structured environment}
\author{Paula I. Villar}
\affiliation{Departamento de F\'\i sica {\it Juan Jos\'e
Giambiagi}, FCEyN UBA and IFIBA CONICET-UBA, Facultad de Ciencias Exactas y Naturales,
Ciudad Universitaria, Pabell\' on I, 1428 Buenos Aires, Argentina.}
\author{Alejandro Soba}
\affiliation{ Centro  At\'omico  Constituyentes,  Comisi\'on  Nacional  de  Energ\'\i a  At\'omica,
Avenida  General  Paz  1499,  San  Mart\'\i n,  Argentina}
\date{\today}                                           

\begin{abstract}
We study the role of driving in an initial maximally entangled state evolving under the presence of a structured environment in a weak and strong regime. We focus on the enhancement and degradation of maximal Concurrence when the system is driven on and out of resonance for a general evolution, as well as the effect of adding a transverse coupling among the particles of the model. We further investigate the role of driving in the acquisition of a geometric phase for the maximally entangled state. As the model studied herein can be used to model experimental situations such as hybrid quantum classical systems feasible with current technologies, this knowledge can aid the search for physical setups that best retain quantum properties under dissipative dynamics. 

\end{abstract}

\maketitle

Two-qubit states are the simplest quantum mechanical systems displaying entanglement. For a bipartite system, it is important to know whether the system is entangled, separable, classically correlated or quantum correlated. As a valuable resource in quantum information processing, entanglement attracts much attention from researchers either in theory or in experiment and much progress concerning this issue has been achieved \cite{Nielsen}.
However, the presence of an environment can destroy all traces of  quantumness of the system. All real world quantum systems interact with their surrounding environment to a greater or lesser extent. As the quantum system is in interaction with an environment, a degradation of pure states into mixtures takes place. No matter how weak the coupling that prevents the system from being isolated, the evolution of an open quantum system is eventually plagued by nonunitary features like decoherence and dissipation. 
Nowadays, decoherence stands as a serious obstacle in quantum information processing. Likewise,
 very interesting effects occur  regarding their entangling properties when two qubits and external environments are considered. For example, two qubits interacting  with two different baths exhibit sudden death of entanglement (ESD) \cite{Eberly}, being more pronounce at finite temperature  \cite{James} and in the presence of external driving \cite{Paranou}. 
This means a major obstacle in building a quantum computer since it can be more difficult to maintain quantum correlations when qubits interact with different reservoirs and are locally driving by single-qubit quantum gates. If qubit-qubit interactions are also considered, another interesting feature of entanglement called ``steady-state generation of entanglement" takes place \cite{Paranou2}.
 The nonintuitive character of entanglement derives from the fact that it affects a bipartite system although it is subject only to local interactions. Entanglement, however, is a characteristic of the whole system, has a dynamics that can not be understood by just adding the sum of the local effects. The qubit-qubit coupling is established either directly, through mutual inductances and capacitances, or as a second-order effect, due to the interaction through a common environment.

The possibility of exploiting the environment as a resource for control has opened a new door in the manipulation
of open quantum systems. The generation and stabilization of entanglement is one of the main challenges for quantum information applications.  The dynamical behavior of correlations in the bipartite system depends on the noise produced by the environment. Many different approaches have studied the entanglement properties of bipartite systems in different frameworks. In \cite{lofranco}, authors have studied quantum correlations in a two-qubit system coupled to either a common bath or two different baths, focusing on environmental memory effects induced on the bipartite system.  In \cite{PRL100}, it has been proposed the used of the Quantum Zeno effect to protect the degradation of the entanglement for  two atoms in a lossy resonator. In \cite{Paranou2},  the interplay of driving and decoherence in a bipartite system under the secular approximation for a dephasing model has been studied. In \cite{majo}, authors numerically studied the generation of a driven-dissipative steady entanglement state for solid state qubits driven by periodic fields, in a dephasing model. Furthermore, in \cite{mirkin} authors focused on driving the open quantum system from an initial separable state to an entangled target state by  exploiting the environment as a resource for control. All of these approaches  had to resort to simplified analytical approaches or  environmental models so as to derive master equations and obtain the dynamical behavior of the bipartite system under particular situations. 

An effective method that avoids approximations was developed by Tanimura \cite{Tanimura} who established a set of hierarchical equations that includes all order of system-bath interactions. System-bath correlations are fully accounted during the time evolution, as compared to the traditional master equation treatments where correlations are truncated at second order. The information concerning the system-bath coherence is stored in the hierarchical elements, which allows us to simulate the quantum entangled dynamics between the system and the environment beyond any analytical approximation. In \cite{PRA85}, authors used the hierarchy equation model to study two noninteracting qubits embedded in a bosonic environment. Therein, they showed the discrepancy between the results obtained in a resonant case with those derived by using the rotating wave approximation for good and bad cavities.
In this manuscript, we extend this model to a driven bipartite system, say  two qubits interacting with a common bath and among them (if desired), driven at and off resonance.   The importance of the driven two-state model is especially pronounced in quantum computation and quantum technologies, where one or more driven qubits constitute the basic building block of quantum logic gates \cite{Nielsen}. Different implementations of qubits for quantum logic
gates are subjected to different types of environmental noise, i.e., to different environmental spectra.
 The dynamics will depend on three ingredients: driving, coupling and dissipation. It is then worth asking if the driving might play any role in the preservation of quantum correlations and under what initial conditions.   Herein, we shall consider a structured environment and have no limitations on the coupling. We can model weak or strong couplings since we are using a numerical method that contemplates the full dynamics of the system, including both dissipation and Lamb shift, which are fully contained in the numerical model. The reason for choosing this model is twofold. On the one side, it is a numerical model that allows as to study the complete environmental induced dynamics of the bipartite system, stressing under what conditions external driving can enhance (or no) quantum correlations of the system for an initial MES. The model accounts for external driving on the particles, dipolar interaction with the environment and transverse coupling among the particles. 
On the other, it is the natural sequel to previous studies of the geometric phase acquired by a driven two level system \cite{PRAdriven} and undriven bipartite system in a spin boson model \cite{bipartite, annals}. This manuscript constitutes a  study of quantum correlations and its preservation for a driven bipartite system in a structured environment, and the further consequences on  the geometric phase. We focus on the preservation of quantum correlations under a nonunitary evolution of initially maximally entangled states (MES) since they are said to posses a robustness condition. This papers is organized as follows: in Section \ref{model}, we present the model consisting of a bipartite two-level system described by a time-periodic hamiltonian interacting with a structured environment, allowing mutual interaction among the particles if necessary. In Section \ref{dynamics}  we numerically solve the dynamics of the system for a weak and strong  evolution through the hierarchy method beyond the rotating-wave approximation for different class ``X" of bipartite state. In the case of one excitation present in the system, we focus the study on what type of conditions (say the initial state, resonance condition and the external driving) prolong or degrade the quantum entanglement compared to the undriven situation. When considering two excitations present in the system we study the geometric phase acquired by the bipartite and compare the results obtained to previous one for the undriven case in a dephasing model. Finally, in Sec. \ref{conclusions} we summarize the results and present conclusions.


\section{The Model}
\label{model}

We consider two qubits independently coupled to a structured environment. The total Hamiltonian which describes this model reads (we set $\hbar=1$ from here on):
\begin{equation}
\hat{H}=\hat{H}_S(t) + \hat{H}_I+ \hat{H}_{\cal E}, \nonumber
\end{equation}
with
\begin{eqnarray}
\hat{H}_S &=&  \bar{\omega}_1 (t) \sigma_+^1\sigma_-^1 \otimes \mathbb{1}_2 + \bar{\omega}_2 (t)\mathbb{1}_1 \otimes  \sigma_+^2 \sigma_-^2 \nonumber \\
&+& \frac{J}{2}(\sigma_+^1\otimes  \sigma_-^2 + \sigma_-^1 \otimes \sigma_+^2)  \\
\hat{H}_I &=& (\sigma_x^1\otimes \mathbb{1}_2 + \mathbb{1}_1 \otimes  \sigma_x^2) \sum_k  (g_k b_k + g_k^*b_k^{\dagger})   \\
\hat{H}_{\cal E} &=& \sum_k \bar{\omega}_k b_k^{\dagger} b_k,
 \label{H}
\end{eqnarray}
where  $\sigma_{\pm}^j= \sigma_x^j \pm i \sigma_y^j$ (with $\sigma_{\alpha}^j$ ($\alpha=x,y,z$) the Pauli matrices for each particle's subspace, i.e $j=1,2$ ) and $b_k$, $b_k^{\dagger}$ the annihilation and creation operators corresponding to the $k-$th mode of the bath. In the $\hat{H}_S$ we are considering a parameter $J$ to include (or not) a transverse coupling among qubits (that can be associated to the mutual capacitance when dealing with coupled flux qubits \cite{majo,wendin}). 
The environmental coupling constant is $g_k$ and $\bar{\omega}_1(t)$, $\bar{\omega}_2(t)$ are the time-dependent frequencies. We shall assume they have the following form:
\begin{eqnarray}
\bar{\omega}_1(t)&=& \bar{\Omega}_1 + \bar{\Delta}_1 \cos({\bar {\omega}_{D_1}} t + \varphi_1),~~~~{\rm and} \nonumber \\
\bar{\omega}_2(t)&=& \bar{\Omega}_2 + \bar{\Delta}_2 \cos({\bar {\omega}_{D_2}} t +\varphi_2)
\end{eqnarray}
where an arbitrary driving field is applied over each atom 1 or 2, with driving frequency $\bar{\omega}_D^i$ (allowing time-dependent energy difference between states $|0\rangle$ and $|1\rangle$ of each two-level particle).
The exact dynamics of the system in the interaction picture has been derived as in \cite{Tanimura, PRA85}. If the qubits and the bath are initially in a separable state, i.e. $\rho(0)=\rho_s(0)\otimes_k |0_k \rangle$, the formal solution is:
\begin{eqnarray}
\tilde{\rho}_S(t) &=& {\cal T} \exp\bigg\{-\int_0^t dt_2\int_0^{t_2} dt_1 \tilde{V}(t_2)^\times  \\
&& [C^R(t_2-t_1)\tilde{V}(t_1)^{\times} + i C^I(t_2-t_1)\tilde{V}(t_1)^{\circ}] \bigg\}\rho_s(0), \nonumber
\label{rhos}
\end{eqnarray}
where  ${\cal T}$ implies the chronological time-ordering operator and we have introduced the following notation $A ^{\times} B=[A,B]= AB-BA$ and $A^{\circ}B= \{A,B\}= A B+ B A$. In our particular case, the interaction potential is a dipolar one  defined by $V= (\sigma_x^1\otimes \mathbb{1}_2 + \mathbb{1}_1 \otimes \sigma_x^2) $ as can be seen in the definition of the Hamiltonian Eq. (2). The coefficients
$C^R(t_2-t_1)$ and $C^I(t_2-t_1)$ in Eq. (5) are the real and imaginary parts of the bath time-correlation function, defined as
\begin{eqnarray}
C(t_2-t_1) &\equiv& \langle B(t_2) B(t_1) \rangle = {\rm Tr}[B(t_2)B(t_1)\rho_B] \nonumber \\
&=& \int_0^{\infty} d\omega J(\omega) e^{-i \omega (t_2-t_1)}
\end{eqnarray} 
and 
\begin{equation}
B(t)=\sum_k \bigg(g_k b_k \exp(-i \omega_k t) + g_k^*b_k^{\dagger} \exp(i \omega_k t)\bigg). \nonumber 
\end{equation}
The difficulty in solving the time ordered integral in Eq. (5) has been overcome by the hierarchy method derived  in \cite{Tanimura, Sun}.  The key condition in deriving the hierarchy equations is that the correlation function can be decomposed into a sum of exponential functions of time.
At finite temperatures, the system-bath coupling can be described by the Drude spectrum. However, if we consider qubits devices, they are generally prepared in nearly zero temperatures.  In such cases of cavity-qubits systems, the coupling spectrum is usually a Lorentz type defined as $J(\omega)$ as has been explained in \cite{Sun2, PRA85},
\begin{equation}
J(\omega)= \frac{{\bar{\gamma}_0}}{2 \pi} \frac{\lambda^2}{(\omega- \bar{\Omega}_0)^2 + \lambda^2}. \label{densidad}
\end{equation}
In the above equation, $\bar{\Omega}_0$ is considered as the average value among the frequencies of the system at $t=0$, $\bar{\gamma}_0$ represents the coupling strength between the system and the bath and 
$\lambda$ characterizes the broadening of the spectral peak, which is connected to the bath correlation time $t_c= \lambda^{-1}$. The relaxation timescale  is determined by $t_r=\bar{\gamma}_0^{-1}$. Then, if we consider the bath in a vacuum state at zero temperature, the correlation function can be expressed as
\begin{equation}
C(t_2-t_1)= \frac{\lambda \bar{\gamma}_0}{2} \exp([-(\lambda + i \bar{\Omega}_0)|t_2-t_1|])
\end{equation}
which is the exponential form required for the hierarchy method. 
Therefore, we can study the
full spectrum of behavior by solving the hierarchy method, which can be expressed as 
\begin{widetext}
\begin{equation}
\frac{d}{d\tau}\rho_{\vec{n}}(\tau)= -(i H_s[\tau]^{\times}+ \vec{n}.\vec{\nu})\rho_{\vec{n}}(\tau) - i \sum_{k=1}^2 V^{\times}\rho_{\vec{n}+\vec{e}_k}(\tau)- i\frac{{\gamma_0}}{2}
\sum_{k=1}^2 n_k [V^{\times} + (-1)^k V^{\circ}] \rho_{\vec{n}-\vec{e}_k}(\tau),
\label{hierarchy}
\end{equation}
\end{widetext}
where we have defined dimensionless parameters variables $\tau=\lambda t$ and $x=\bar{x}/\lambda$ (where $x$ is any parameter with units of energy in the model described).
The subscript $\vec{n}=(n_1,n_2)$ with integers numbers $n_{1(2)} \geq 0$, and $\rho_S(t) \equiv \rho_{(0,0)} (t)$. 
We want to emphasize that  the ``physical" solution is encoded in $\rho_{(0,0)} (t)$ and all other $\rho_{\vec{n}}(\tau)$ with $\vec{n} \neq (0,0)$ are auxiliary operators, just defined in order to solve the system.
The vectors $\vec{e}_1=(1,0)$, $\vec{e}_2=(0,1)$ and $\vec{\nu}=(\nu_1,\nu_2)=(1-i \Omega_0, 1 + i \Omega_0)$.
This set of linear differential equations can be solved by the use of a Runge Kutta routine. It is important to mention that for numerical computations, the hierarchy equations
must be truncated for large $\vec{n}$. The hierarchy terminator equation is similar to that of Eq. (\ref{hierarchy}) for the term $\vec{N}$, and the corresponding term related to $\rho_{\vec{N}+\vec{e}_k}$ is dropped \cite{Tanimura}. In all simulations presented in this manuscript, we have set the order of truncation at $\vec{N}=(20,20)$, as we have checked the convergence of positive reduced matrix $\rho(\tau)$.
This method can describe the dynamics of a system with a nonperturbative and non-markovian system-bath interaction at finite temperature, even under strong time-dependent perturbations. This formalism is valuable because it can be used to study not only strong-bath coupling, but also quantum coherence or quantum entanglement. The information concerning the system-bath coherence is stored in the hierarchical elements, which allows us to simulate the quantum entangled dynamics between the system and the environment.
In the particular case of this manuscript, we shall focus on the entangled dynamics of a driven bipartite system and its environment.

\section{Environmentally induced Dynamics on maximally entangled bipartite states}
\label{dynamics}

For the particular purpose of this work, we shall  consider entangled two-qubits states within the class of X-states, defined in the computational standard basis ${\cal B}=\{|1> \equiv |11>, |2> \equiv |10>, |3> \equiv |01>, |4> \equiv |00> \}$\cite{lofranco}.
\begin{equation}
\rho_X=
\begin{pmatrix}
\rho_{11} & 0 & 0 & \rho_{14}\\
0 & \rho_{22} & \rho_{23} & 0 \\
0 & \rho_{23}^* & \rho_{33} & 0\\
\rho_{14}^* & 0 & 0 & \rho_{44}
\end{pmatrix}.
\label{rho1}
\end{equation}
This type of matrices are commonly used in a variety of physical situations. For several physical dynamics this structure is further maintained in time. These X states contain the Werner-like states defined as
\begin{equation}
\rho_r(0)=\frac{1-r}{4}\mathbb{1} + r|\tilde{\phi}\rangle \langle \tilde{\phi} \rangle
\label{rhoi}
\end{equation}
where $r \in (0, 1]$ determines the mixing of the state and $\mathbb{1}$ is the unit matrix in the Hilbert space $4 \times 4$. The state $|\tilde{\phi}\rangle$ may be any of the following states:
\begin{eqnarray}
|\Phi_{\pm} \rangle &=&\sqrt{1-p}|01\rangle \pm \sqrt{p}|10\rangle, \\  
|\Psi_{\pm} \rangle &=& \sqrt{1-p}|00\rangle \pm \sqrt{p}|11\rangle 
\label{Bell}
\end{eqnarray}
where $p$ determines the degree of entanglement, being $|0\rangle$ and $|1\rangle$
eigenstates of the Pauli operator $\sigma_z$. It is easy to note that
when $p = 1/2$, the above states are the Bell states. For $r=0$ the Werner-like states become totally mixed states, while for $r=1$ they reduce to the Bell states in Eqs. (12) and (\ref{Bell}). 
There are two subclasses, namely states that are diagonal on a subspace of single excitations and states that are diagonal in the subspace of zero and two excitations. In the following, we shall consider an initial bipartite state of each of the two subclasses and study the Concurrence as it evolves in time and driving is included in the system. It is important to recall that we shall work with dimensionless frequencies and temporal parameter $\tau$, say $\Omega_i=\bar{\Omega}_i/{\lambda}$, $\Delta_i=\bar{\Delta}_i/\lambda$, $\omega_D ^ i=\bar{\omega}_D^i/\lambda$.  We shall assume a bath correlation time $\tau_c=1$.

\subsection{Initial maximally entangled states $|\Phi_{\pm}\rangle$}

In this subsection, we shall focus on the  main features of the driven non-interacting qubits when embedded  in a common structured environment.  We shall consider only one excitation is present in the system and the environment is initially in vacuum. This means that at $t=0$, the cavity is initially in a vacuum state $|0_k \rangle$. We consider the initial state as
\begin{equation}
|\Phi_+\rangle= (\sqrt{1-p}|01\rangle + \sqrt{p}|10\rangle) \otimes_k |0_k\rangle
\end{equation}
which means by defining the initial density matrix in Eq. (\ref{rhoi}) with $r=1$ and $|\tilde \phi\rangle=|\Phi_+\rangle$.  The 
reduced density matrix takes the form
\begin{equation}
\rho(t)=
\begin{pmatrix}
 0& 0 & 0 & 0\\
0 & \rho_{22}(t) & \rho_{23}(t) & 0 \\
0 & \rho_{23}^*(t) & \rho_{33}(t) & 0\\
0 & 0 & 0 & 1- \rho_{22}(t)- \rho_{33}(t)
\end{pmatrix}.
\label{rho2}
\end{equation}

\begin{figure}[h] 
	\includegraphics[width=8.cm]{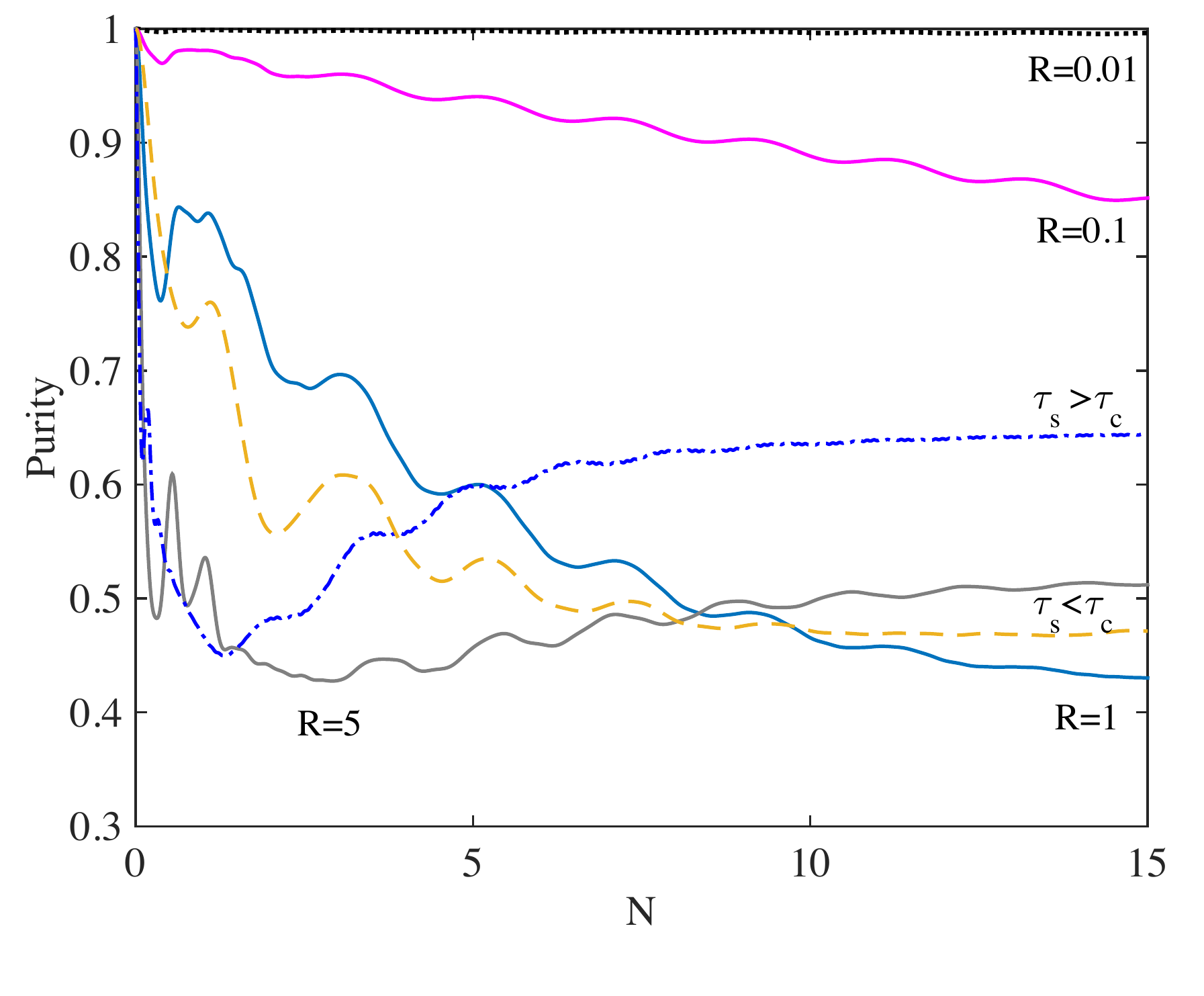}
 	\caption{(Color online) We show the degradation of Purity of the bipartite system as time evolves (measured in  natural cycles $N=\tau/\tau_s$)  for different couplings to the environment.Solid lines represent the same value for the system parameters 
 	 ($\Omega_1=10$,  $\Omega_2=15$) but different strength couplings $R$. For these values we have $\tau_s \sim \tau_c$.
 	The dashed line is a representative value for $\tau_s < \tau_c$ ($\Omega_1=10$, $\Omega_2=20$), with a strong environment $R=1$. Finally, the dot-dashed line is a representative situation for $\tau_s > \tau_c$, for a near-resonance situation with $R=1$:  $\Omega_1=10$, $\Omega_2=11$. 
 	 As it can be seen, a stronger coupling to the environment leads to a bigger loss of Purity ($R=5$). As $\tau_s$ represents proper times, it indicates different timescales in the diverse situations (N=15 for the dashed line represents a shorter time than for the others).  All parameters are dimensionless ($\Omega_i/\lambda, ~\Delta_i/\lambda$), $\tau_c=1$, $R=\gamma_0/\lambda$, $J=0$. No driven is considered.}
 	\label{Fig1}
 \end{figure}

We can start by studying the loss of Purity as time evolves in the case of an initial maximally entangled state (MES), say $p=1/2$. The dynamics of the system is determined by the coupling between the system and the environment. It is commonly found in the Literature, the definition of bad and good cavity determined by the value of the factor $R=\bar{\gamma}_0/\lambda$. This means that for a weak coupling, we have $\bar{\gamma}_0 < \lambda$, leading to a $R<1$, commonly known as bad cavity limit. Otherwise, for a stronger coupling, $R>1$ yields a good cavity limit. We shall use this convention to characterize the strength of the coupling. These qualitatively different behaviors are shown with solid lines in Figure \ref{Fig1}. 

  Therein, we can see that by changing the value of the coupling constant $\bar{\gamma}_0$, we can simulate different cavities behaviors (leaving $\lambda$ fixed). For strong coupling we can see the evolution is accompanied by the typical fast oscillations of non-markovian evolutions (solid lines for $R=1$ and $R=5$).
In Figure \ref{Fig1}, we  have also included  weak coupling ($R=0.01$ and $R=0.1$) for the same dynamical timescale $\omega_s=\omega_2-\omega_1$ and $\tau_s=2\pi/\omega_s$. Further,  we have added  two other representative situations for a strong coupling $R=1$. The dot-dashed line is for a quasi-resonant case, meaning both particles have very similar frequencies. This yields  $\tau_s>\tau_c$. On the contrary, the dashed line is for two considerably different atom frequencies with $\tau_s < \tau_c$. As $N$ indicates the number of natural cycles elapsed, then it is instructive to recall that it represents longer times for longer $\tau_s$. We can note that the loss of Purity is done in a few natural cycles in the near-resonance case, for a strong environment $R=1$ (even faster than for the case $R=5$ and $\tau_s \sim \tau_c$). This means that, depending on the system's timescale, the transition between atomic states has already been achieved for $N=15$, leading to an increment in the population of $|00\rangle$ state ($\rho_{00}(t)=1-\rho_{22}(t)-\rho_{33}(t)$, see Eq.(\ref{rho2})), and therefore of Purity (see appendix Section \ref{appendixa}). Likewise, for a shorter $\tau_s$, it seems that a stationary state has not already been achieved.
The advantage of this numerical method is that  by choosing the right set of parameters, we can simulate a different type of environment and obtain the corresponding dynamics beyond the rotating-wave approximation.

In the following,  we shall focus on how  external driving can affect (or enhance) the entanglement dynamics under a strong regime $R=1$. We shall consider the qubits to have similar (but not equal) frequencies yielding  $\tau_s \sim \tau_c$.  
In particular, we shall investigate to what extent external driving  acting solely on the bipartite system can preserve quantum correlations in a strong regime  where memory effects of the environment take place.  
The dynamics of the bipartite two-level driven system comprises three different dynamical effects, each occurring on a different timescale. Dissipation and decoherence occur on the relaxation timescale $\tau_r$ and non-Markovian memory effects occur for times shorter than or similar to the reservoir correlation timescale $\tau_c$ \cite{PRAdriven} (in addition to the system's timescale $\tau_s$). 
 
Quantum decoherence implies a rapid reduction of the off-diagonal terms of the bipartite reduced density matrix. The quantity for measuring the entanglement between the different parts of the composite system is the Concurrence \cite{Eberly, lofranco}. The Concurrence for the evolution of this state can be computed as ${\cal C}(\tilde{\rho}_r)=\rm{max}(0, \sqrt{\lambda_1}-\sqrt{\lambda_2}-\sqrt{\lambda_3}-\sqrt{\lambda_4} )$, where $\lambda_1$, $\lambda_2$, $\lambda_3$ and $\lambda_4$ are the eigenvalues of $\tilde{\rho}_r = \rho_r(\sigma_y^1\otimes \sigma_y^2) \rho_r ^*(\sigma_y^1\otimes \sigma_y^2)$.
 \begin{figure}[h!]
	\includegraphics[width=7.5cm]{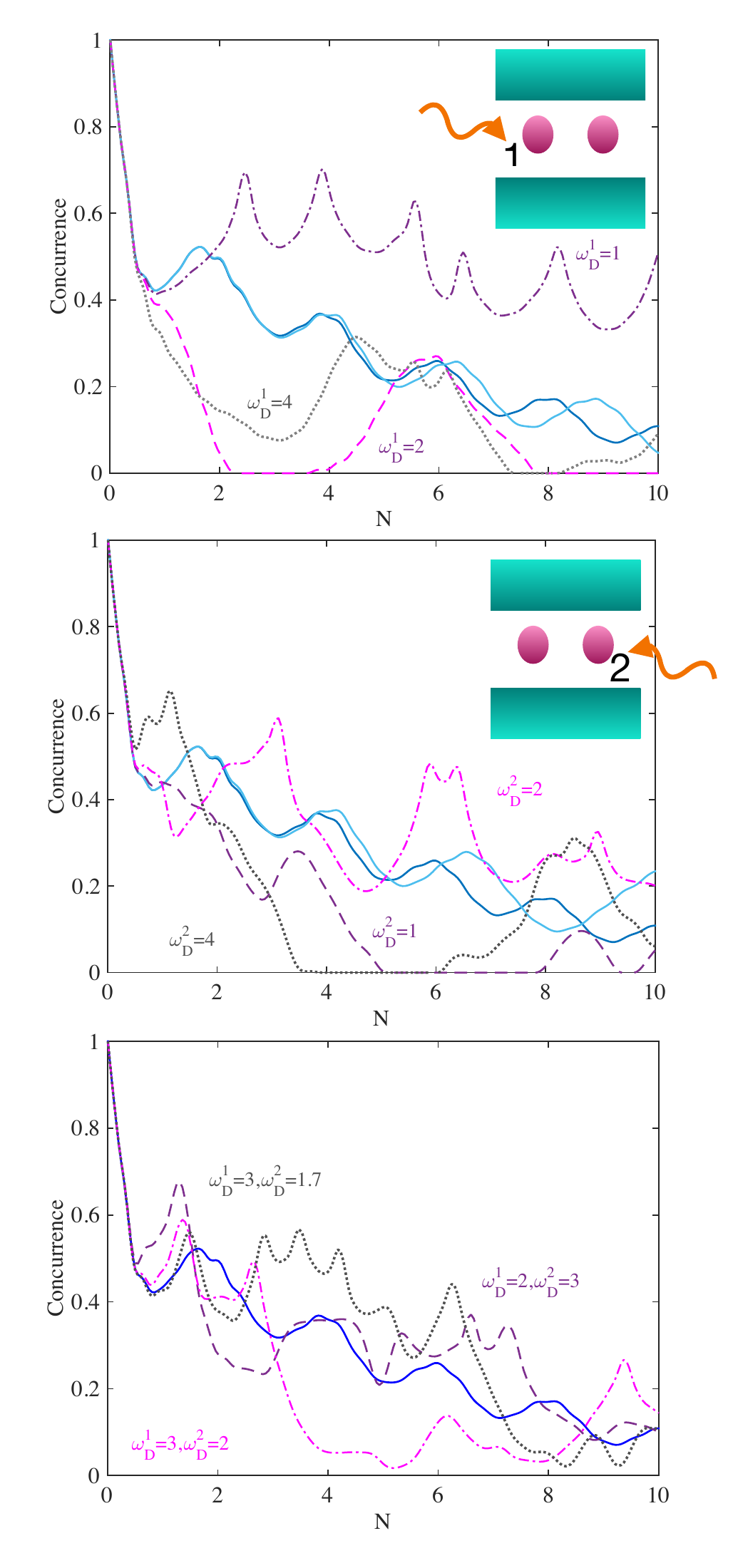}
 	\caption{(Color online) We show how the external driving on each particle affect the Concurrence of the bipartite initial system as time evolves (measured in  natural cycles $N$).  At the top panel, we add driving frequency to the one of the particles composing the bipartite system ($\omega_D^1=0.1$, $\omega_D^1=1$, $\omega_D^1=2$ and $\omega_D^1=4$). In the middle panel we add driving to the other $\omega_D^2=0.1$, $\omega_D^2=1$, $\omega_D^2=2$ and $\omega_D^2=4$.   Finally, in the bottom panel, we add driving to both particles ($\omega_D^1=3$ and $\omega_D^2=1.7$, $\omega_D^1=2$ and $\omega_D^2=3$ and $\omega_D^1=3$ and $\omega_D^2=2$).  Parameters: $R=1$, $\tau_c=1$, $\Omega_1=15$, $\Omega_2=10$, $\Delta_1=4$, $\Delta_2=7$, $\varphi_1=\pi$,  $\tau_s=1.04$, $J=0$. }
 	\label{Fig2}
 \end{figure}
We start by adding driving to only one of the particles, assuming similar frequencies $\Omega_i$ and different detuning ones $\Delta_i$.
In Figure \ref{Fig2}, we show the behavior of the Concurrence as time evolves for different driving situations. In the top panel of Figure \ref{Fig2} we add driving to one of the particles, say particle number 1. Then, $\omega_1(t)=\Omega_1 + \Delta_1 \cos(\omega_D^1 t +\varphi_1)$, and leave the other particle undriven $\omega_2=\Omega_2$. We can see the behavior is qualitatively different as the driving frequency varies. The blue solid line is for $\omega_D^1=0$, which means no driving at all, and can be used as a reference for the undriven case. The light blue dashed line is for $\omega_D^1=0.1$ and the purple dotted-dashed line for $\omega_D^1=1$. The magenta dashed line represents $\omega_D^1=2$ and the gray dotted line $\omega_D^1=4$ (all frequencies have been adimensionalized by $\omega_D=\bar{\omega}_D/{\lambda}$).  We can even note that for  $\omega_D^1=2$ and $\omega_D^1=4$ there is entanglement sudden death (ESD) occurring at different times, while for $\omega_D^1=1$, Concurrence is enhanced when compared to the static situation. 

 We can also study the behavior of Concurrence as the driven particle is number 2 with  $\omega_2=\Omega_2 + \Delta_2 \cos(\omega_D^2 t+ \varphi_2)$ and number 1 is static (middle panel). The blue solid line is for $\omega_D^2=0$ and light blue solid line for $\omega_D^2=0.1$. We can see that Concurrence has bigger values at longer times when driven at a small driving frequency (even compared to the specular situation in top panel). The purple dashed line represents $\omega_D^2=1$, the magenta dot-dashed line  $\omega_D^2=2$ and the gray dotted line $\omega_D^2=4$ .
 It is interesting to note the qualitatively different behaviors obtained for similar values in both different panels. Finally, we can add driving  to both particles at the same time. The behavior of the Concurrence is shown in the bottom panel for some set of values of $\omega_D^1$, $\omega_D^2$. As can be seen, it is not that easy to predict the behavior of the Concurrence when both particles are driven. However, we can note that for some set of values, Concurrence is enhanced and quantum correlations are evidently preserved over longer times compared to the static case. This result agrees with that found for a single driven atom in \cite{PRAdriven}. 
 In order to show the full picture, we include all set of values for $\omega_D^1$ and $\omega_D^2$ in Figure \ref{Fig3} (a). Therein, we show how driving on both particles affects the evolution of the initial maximum Concurrence for a strong coupling  regime ($R=1$). On the top left corner, we show the Concurrence for the bipartite system having elapsed one natural cycle, say $N=1$. As the interaction with the environment is switch on at $t=0$, Concurrence suffers an abrupt decrease but starts oscillating thereafter. However, the later evolution strongly depends on the external driving, because we can see that we can find areas with greater or lesser degree of entanglement (compared to the static situation) at $N=3$. There are areas where entanglement has increased compared to the similar static situation where driving is not considered (in the bottom left corner). This fact is accented for later times as  can be observed  for $N=5$. Black areas show that entanglement has been abruptly destroyed in agreement with Figure \ref{Fig2}, bottom panel. Concurrence for $N=7$ is shown in the right bottom picture. Therein, we can see that Concurrence is generally destroyed for this timescale in the strong coupling regime for this initial bipartite state. 

\begin{widetext}
\begin{center}
\begin{figure}[h]
	\includegraphics[width=16cm]{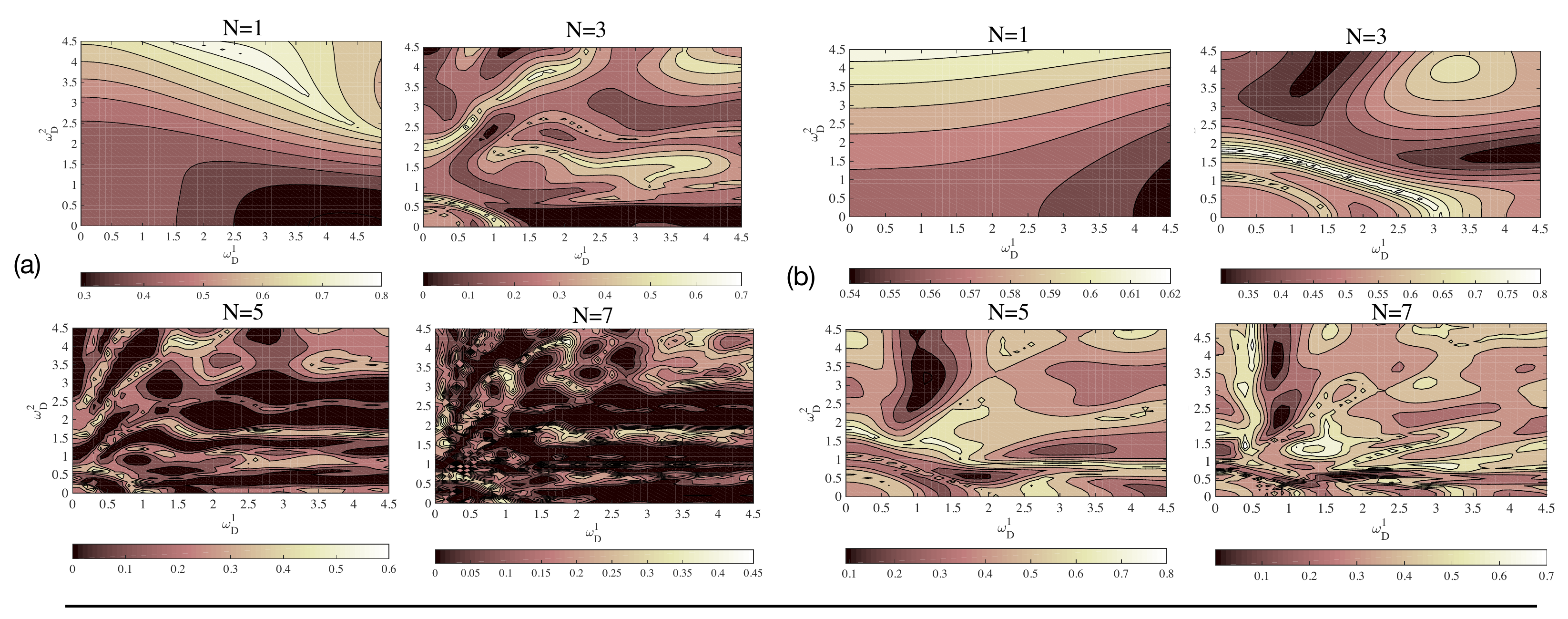}
\caption{(Color online) (a) We show how the external driving on each particle affects the Concurrence of the maximally entangled initial state of the bipartite system for different times (measured in natural cycles $N$) for  different detuning frequencies and  strong coupling.  Parameters used: $R=1$, $\tau_c=1$, $\tau_s=1.04$, $\Omega_1=15$, $\Omega_2=10$,  $\Delta_1=4$, $\Delta_2=7$, $\varphi_1=\pi$ and $J=0$. (b) Concurrence of the bipartite initial system for different times ($N$) for similar detuning frequencies  under strong coupling.  Parameters used: $R=1$, $\tau_c=1$, $\tau_S=1.03$, $\Omega_1=15=\Omega_2$, $\Delta_1=3$, $\Delta_2=3.1$, $\varphi_1=\pi$,  and $J=0$. }
 	\label{Fig3}
\end{figure}
 \end{center}
 \end{widetext}

We can try to understand the results by assuming they depend  on the form of the external driving we are introducing as it moulds the response of the bipartite system against the environment.  By looking at the figure, it seems that higher frequencies imply a constant flux of driving which in turn can be associated with a stronger Concurrence for the initial MES state. On the contrary, if the driving has a  spaced effect in time, its competition against the memory effects of the bath is weak and it may result in a lesser degree of entanglement as time evolves.
It is important to recall that in Figure \ref{Fig3}(a), $\Delta_1 < \Delta_2$ and this fact can lead to biased conclusions. Hence, in order  to see to what extent our results depend on  the value of the detuning frequency, we shall next consider the evolution in time of an initial  maximally entangled state $|\Phi_+\rangle$, under similar detuning frequencies leading to  a quasi-resonant condition. Then, we shall drive the system with $\Delta_1 \sim \Delta_2$, leaving fixed $\tau_s \sim \tau_c$ for a proper comparison.  
In Figure \ref{Fig3}(b) we can observe the corresponding panels for the time evolution of the Concurrence  ($N=1$, $N=3$, $N=5$ and $N=7$).  In this case, patterns seem to be symmetric and quantum correlations seem to prevail for longer times. In $N=5$ there are less regions of null Concurrence than in the corresponding previous case.  In this case, when frequencies are similar, beats take place and their periods are longer, which seems to enhance quantum correlations. For $N=7$ we can find clean areas where quantum correlations are still alive and of greater strength when compared to the corresponding panel of Figure \ref{Fig3}(a).

So far, we have seen that having a highly detuned scenario does not necessarily enhance quantum correlations (Figure 3.(a)). We can see that they disappear soon enough, in some areas even earlier than for the undriven case (left corner of the plots). We believe the non-Markovian environment rules the evolution of the system in this particular case.
However, when the system has similar detuning frequencies $\Delta_i$ (Figure 3.(b)), the dynamics is considerably different, exhibiting regions where quantum correlations are preserved longer compared to the static case.
We can interpret this result by considering the unitary phase factor $e^{-i \int_0^{\tau_s} (\omega_2(t')-\omega_1(t'))dt'}$ in the case
 $\Delta_1 \sim \Delta_2$. In such a situation, the phase factor can be approximated by $ (\Omega_2-\Omega_1) \tau_s+\Delta/\omega_D \sin( \omega_D^R \tau_s)$.  We must note that we have further considered $\tau_s\sim 1$. This means that  driving is the resultant of some other periodic function of period $T_R= 2\pi /\omega_D^R$. 
 This would be equivalent to having a particle of renormalized frequency $\Omega_2-\Omega_1$ to which driving is added. We can interpret the results obtained for this situation by comparing to those obtained for a driven two-level particle in the presence of a structured environment.
 In \cite{Poggi}, authors have studied to what extent the effect of adding driving on the system can increase the non-Markovianity (NM) with respect to the undriven case. They have shown that driving can not increase the degree of NM for strong couplings. Further, authors have shown that there is a large area where NM is suppressed in intermediate couplings ($\bar{\gamma}_0 \sim \lambda$, herein $R=1$) for $\omega_D/\Delta \simeq 1$. It has later been shown that this suppression of NM can have further detectable effects as for example in the geometric phase acquired by a two-level system evolving coupled to a strong environment. It has been evidenced that, when $\bar{\gamma}_0 \sim \lambda$, a two-level driven evolution with $\omega_D/\Delta \sim 1$, verifies the suppression of revivals and ensures a smooth evolution, allowing an acquisition of a geometric phase more similar to the unitary one \cite{PRAdriven}.
 Therefore, we believe that this reported  suppression of NM can be responsible for a preservation of quantum correlations as can be seen in the upper corner of Figure 3.(b) where $\omega_D/\Delta \simeq 1$.
 
 \begin{figure}[h]
	\includegraphics[width=8.cm]{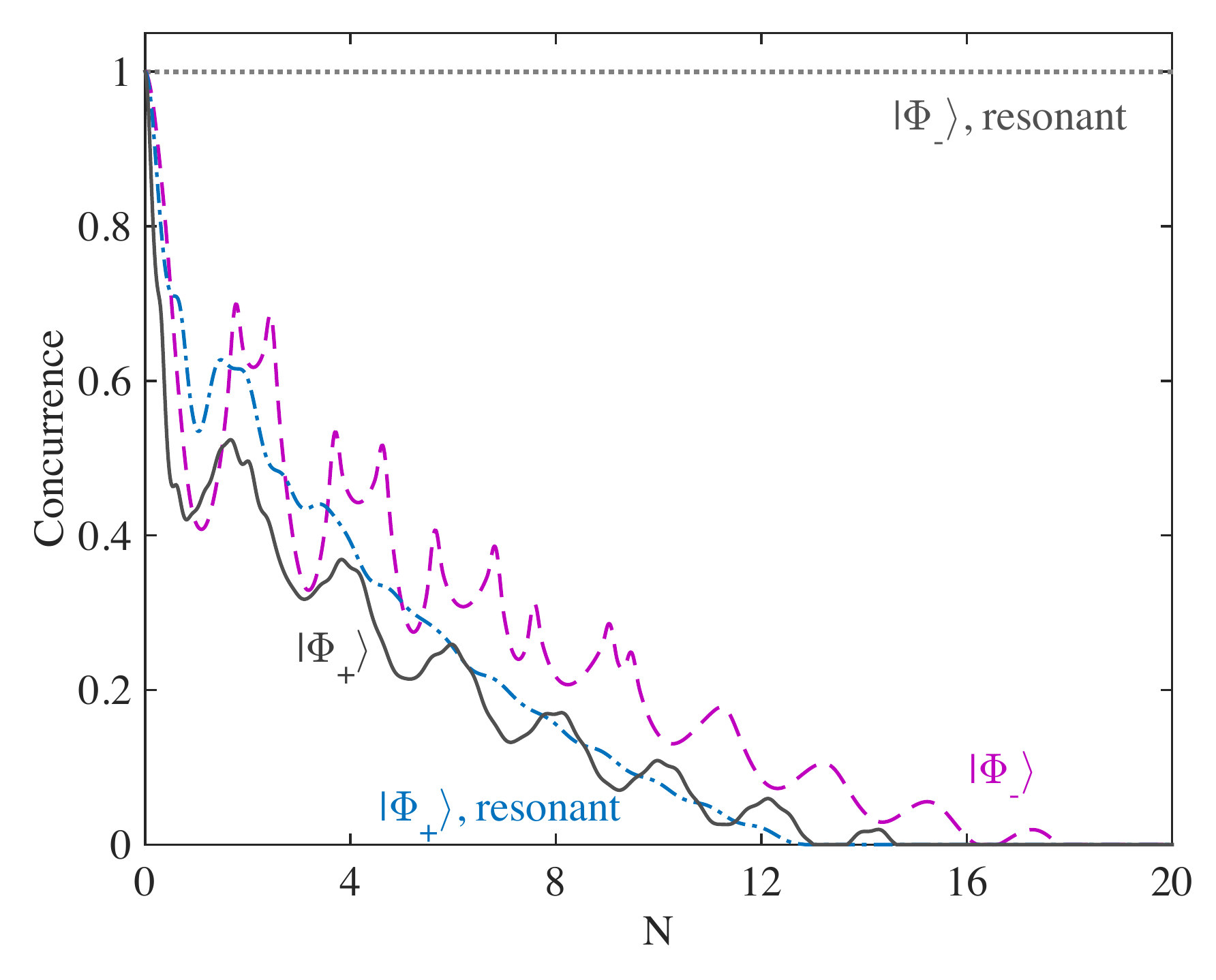}
 	\caption{(Color online). Concurrence as time evolves for initial different maximally entangled stated $|\Phi_+\rangle$ and $|\Phi_-\rangle$, in resonance ($\Omega_1=\Omega_2=10$)  and out of resonance ($\Omega_2=15$ and $\Omega_1=10$). Gray dotted line represents the resonant case of $|\Phi_-\rangle$, while the magenta dashed line is the off-resonance case of the same MES. Dotted-dashed blue line is the resonant case of $|\Phi_+\rangle$, while the dark solid line corresponds to the off-resonant situation of that MES. There is no driving considered.  Parameters: $R=1$, $\tau_c=1$, $J=0$. }
 	\label{Fig4}
 \end{figure}
 
A very interesting result reported in \cite{PRA85, PRL100}, is the steady-state entanglement. It is easy to prove that $|\Phi_-\rangle$  is an eigenstate of $H_S$ and $H_I$ for the particular situation that $\omega_1=\omega_2$ and $J=0$.  
In Figure \ref{Fig4} we show the evolution of  Concurrence for different initial maximally entangled states, $|\Phi_{\pm}\rangle$, in and out of resonance ($\Omega_1=\Omega_2$ and $\Omega_1\neq \Omega_2$ correspondingly, $\Delta_i=0$). The black dotted line corresponds to the resonant case of $|\Phi_-\rangle$, while the magenta dashed line is the off-resonant case,  for $\tau_s \sim \tau_c$. The dark gray solid line represents the off-resonant case for $|\Phi_+\rangle$, while the blue dot-dashed  line  is  the  resonant case of the same initial MES \cite{nota}. As it can be noted in the figure,
Concurrence is enhanced for the initial $|\Phi_-\rangle$ state (with respect to $|\Phi_+\rangle$) given the same parameters of the system and when driving is not considered at all. Therefore, in the following, we shall explore the effect on quantum correlations of adding driving to the initial  MES given by $|\Phi_-\rangle$ under a strong coupling.   

In Figure \ref{Fig5} we show the Concurrence for different times elapsed when the maximally entangled state evolves under a structured environment in a strong regime and $\Delta_1<\Delta_2$. We show different time pictures so as to compare with the Concurrence's time evolution of the initial $|\Phi_+\rangle$ state. On the left top  row  we include $N=1$ on the left and $N=3$ on the right, while in the bottom right we present  $N=5$ on the left and $N=7$ on the right. Similarly, in  Figure \ref{Fig5}(b) we include the Concurrence for different times elapsed when the maximally entangled state evolves under a structured environment in a strong regime when the system is driven with similar detuning frequencies,
assuming $\tau_s \sim \tau_c$. On the left top  row  we include $N=1$ on the left and $N=3$ on the right, while in the bottom right we present  $N=5$ on the left and $N=7$ on the right for the correct comparison among the different initial conditions.

 \begin{widetext}
\begin{center}
\begin{figure}[h]
	\includegraphics[width=16cm]{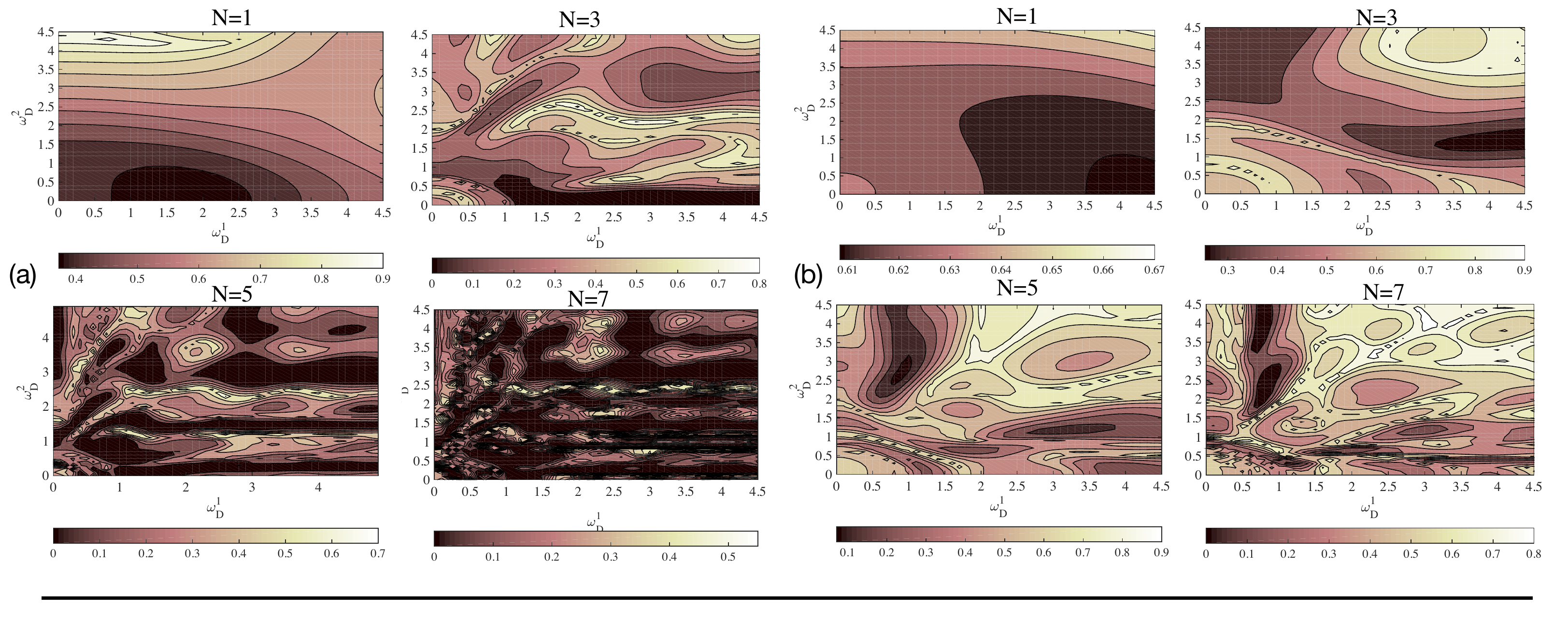}
\caption{(Color online) (a) We show how the external driving on each particle affects the Concurrence of the initial MES  $|\Phi_-\rangle$ of the bipartite system for different times  ($\rm N$) and non similar detuning frequencies under a  strong coupling. Parameters  used: $R=1$, $\tau_c=1$, $\tau_s=1.04$, $\Omega_1=15$, $\Omega_2=10$, $\Delta_1=4$, $\Delta_2=7$, $\varphi_1=\pi$ and $J=0$. (b) Concurrence of the driven bipartite system, by setting the initial MES $|\Phi_-\rangle$, for different times $\rm N$ and similar detuning frequencies  under strong coupling. Parameters used: $R=1$, $\tau_c=1$, $\tau_s=1.03$, $\Omega_1=15=\Omega_2$,  $\Delta_1=3$, $\Delta_2=3.1$, $\varphi_1=\pi$ and $J=0$.}
 	\label{Fig5}
\end{figure}
 \end{center}
 \end{widetext}
 
 \begin{figure}[h]
	\includegraphics[width=8.cm]{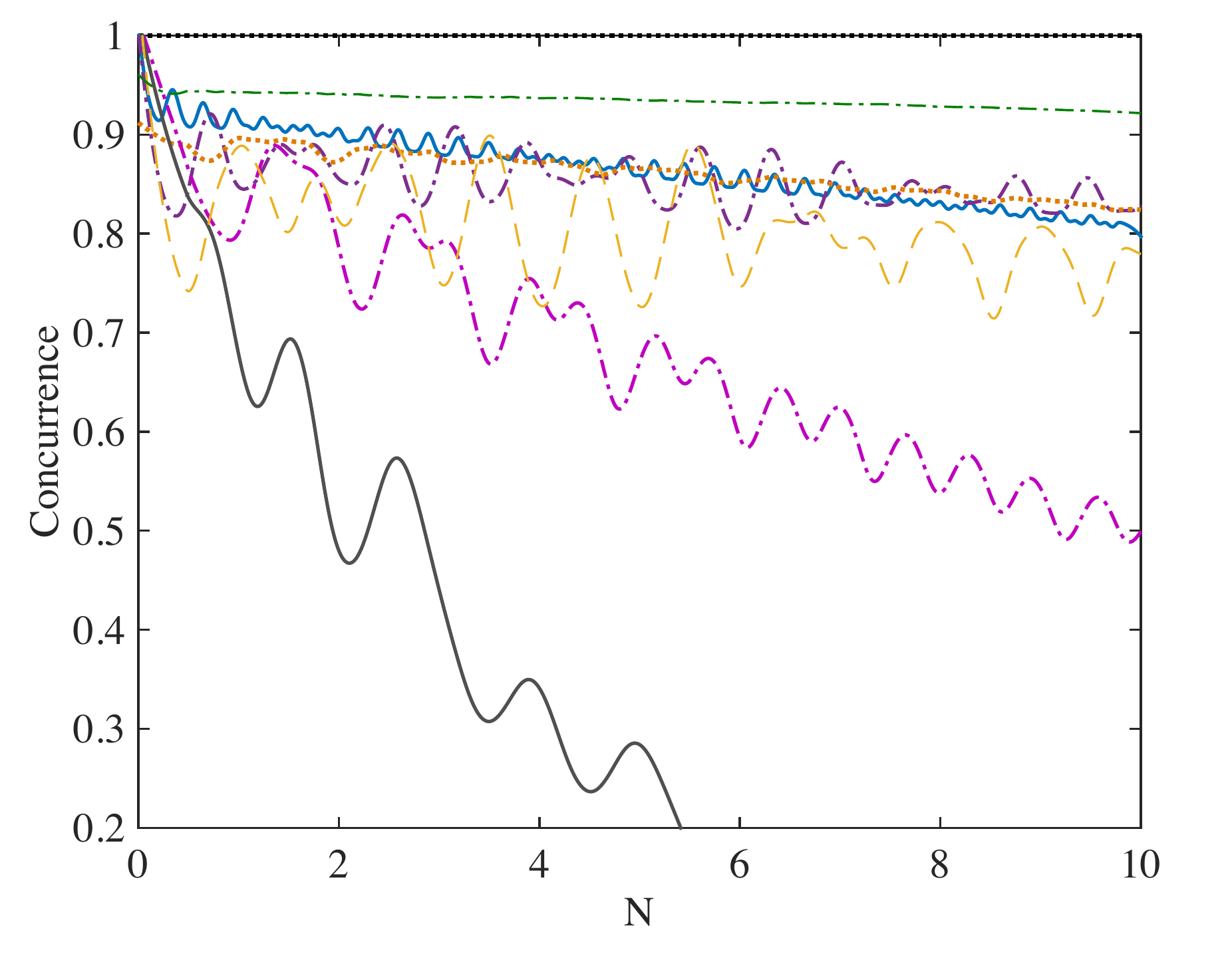}
 	\caption{(Color online) Concurrence as time evolves for an initially MES  $|\Phi_-\rangle$ for different values of $\Delta_1$, $\Delta_2$, $\omega_D^1$ and  $\omega_D^2$. The top dotted black line represents the  undriven case with $\Omega_1=\Omega_2=15$. 
 	The dotdashed green line on top represents a quasi resonant case since $\Omega_1=15=\Omega_2$ and $\Delta_1=0.5$, $\Delta_2=0.5$, $\varphi_1=\pi$, very similar frequencies.
 	The top blue solid  line corresponds to $\Delta_1=1.$, $\Delta_2=1.1$, $\varphi_1=\pi$ and $\omega_D^{1,2}=7$, while  the purple dot-dashed line differs in the driving frequency $\omega_D^{1,2}=3$. The dashed orange line is for $\Delta_1=2.$, $\Delta_2=5$, $\varphi_1=0$ and $\omega_D^{1,2}=3$.
 	The magenta dot-dashed line represents the evolution of $\Delta_1=5.$, $\Delta_2=5.1$, $\varphi_1=\pi$ and $\omega_D^{1,2}=8$. The gray solid line is for  $\Delta_1=4$,          $\Delta_2=7$, $\varphi_1=\pi$,  $\omega_D^{1,2}=5$. The brown dotted line has similar values than the blue solid line but for $J=1$. Parameters of the environment: $R=1$, $\tau_c=1$.}
 	\label{Fig6}
 \end{figure}
 
 When comparing results obtained in Figure \ref{Fig3} and Figure \ref{Fig5}, we can note that similar detuning frequencies exhibits stronger quantum correlation at longer times. Among both cases, we can observe that when the initial MES is $|\Phi_-\rangle$ (Figure \ref{Fig5} (b)), quantum correlations seem to prevail for more set of parameters than when the initial MES is $|\Phi_+\rangle$ (Figure \ref{Fig3} (b)). 
In the particular case of $|\Phi_-\rangle$, for the same timescale, quantum correlations seem to be stronger for some particular values verifying $\omega_D/\Delta \simeq 1$. As in the above figures, the detuning frequencies have been kept fixed,  we shall further explore other different driving situations for the initial state $|\Phi_-\rangle$,  allowing for the existence of interaction among the particles through the parameter $J$ as well.  In Figure 6, we show a dotted black line for the resonant undriven case $\Omega_1=\Omega_2$ as a reference.  
 The blue solid case is for a driven situation of a higher frequency $\omega_D^1=7=\omega_D^2$ than the dot-dashed purple line ($\omega_D^1=3=\omega_D^2$), both lines for $\Delta_1=1$, $\varphi_1=\pi$ and $\Delta_2=1.1$.  We can see that these lines correspond to a near-resonance condition since frequencies are very similar. In this particular case, it is important to note that Concurrence has a small decrease in comparison to the static case.  The green dotdashed line corresponds to a smaller difference among frequencies, a nearer-resonance situation. 
 The orange dashed line represents other values of detuning frequencies and driving: $\Delta_1=2$, $\Delta_2=5$, $\varphi_1=0$ and $\omega_D^{1,2}=3$ (still small but bigger difference than the other cases).
 These values can be well compared to the situation when $\Delta_1$ and $\Delta_2$ are bigger, leading to a greater difference $\omega_2(t)-\omega_1$(t). The dashed magenta line is for $\Delta_1=5$, $\varphi_1=\pi$,  $\Delta_2=5.1$  and $\omega_D^{1,2}=8$. 
The gray solid line is for  $\Delta_1<\Delta_2$, as the driving situations shown in Figure \ref{Fig5}(a). 
These representative values show that a small difference among frequencies $\omega_2(t)-\omega_1(t)$ (hence, a quasi-resonant situation) helps better preserve quantum correlations. Figure \ref{Fig6} adds information to that shown in Figure \ref{Fig5}, because it exhibits the behaviour of different sets of detuning frequencies $\Delta_i$. Finally, in that figure, we have also added a line where we included a transverse interaction among the particles with a brown dotted line $J=1$ (all other parameters have the same values than the blue line). By including this transverse coupling, we further show that our results are in agreement with other previous studies, where authors have shown that transverse coupling is less harmful than longitudinal coupling. 
  All in all, we can state that enhancement of Concurrence occurs near resonant condition for $|\Phi_-\rangle$ state.  Furthermore, we can mention that if we include in the system's Hamiltonian transverse coupling, Concurrence seems to be robust (brown dotted line). This result agrees with that achieved by authors in \cite{majo} where they studied the manipulation of quantum entanglement in coupled flux qubits for transverse and longitudinal coupling in the context of closed systems. In \cite{solid} authors found that transverse coupling was less harmful than longitudinal noise for low and high frequency of external noise in agreement with the observation of the Berry phase in an experimental setup \cite{leek}. 
  \begin{figure}[h]
	\includegraphics[width=7.cm]{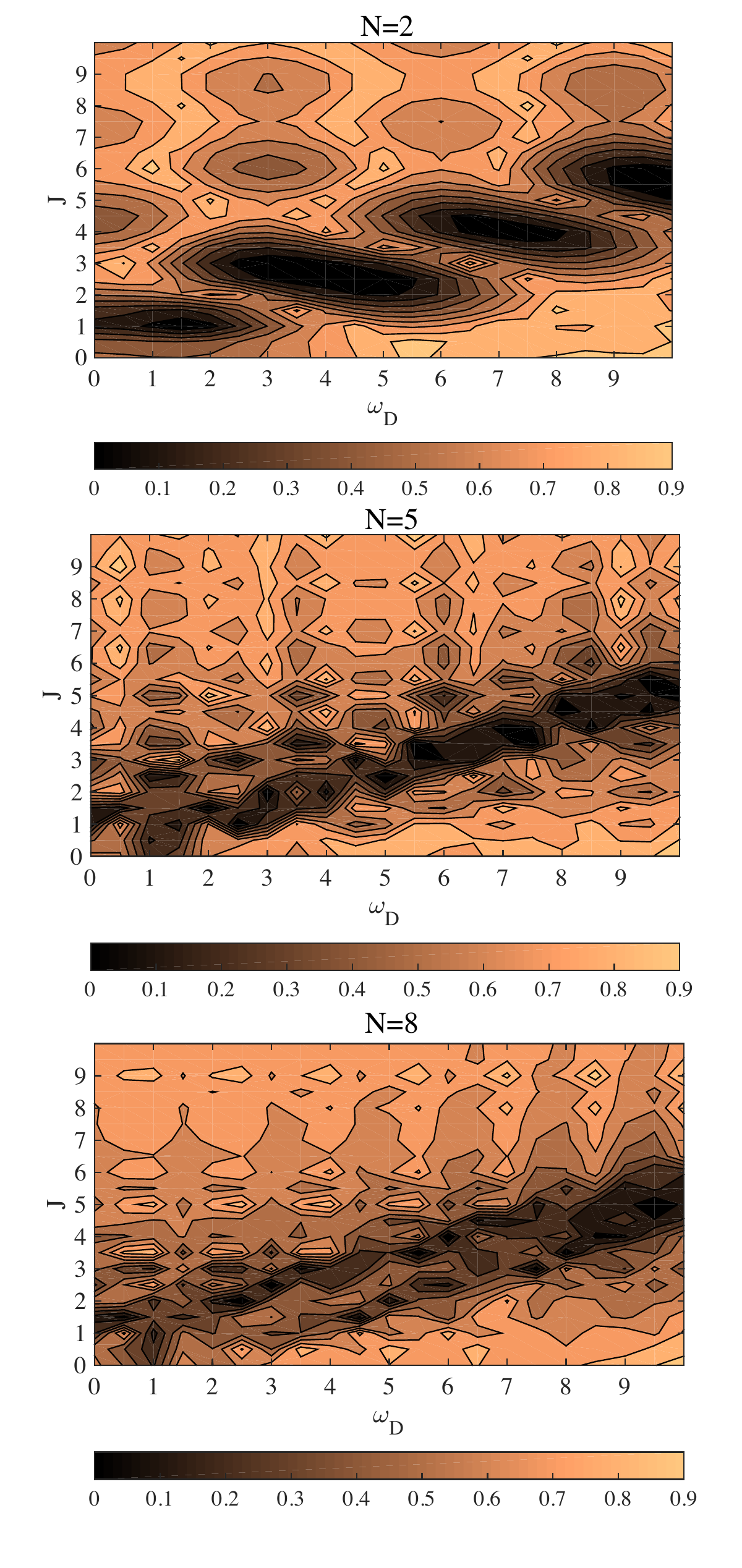}
 	\caption{(Color online) Concurrence as time evolves for an initially MES $|\Phi_-\rangle$ driven to a quasi-resonant case, for different values of  $\omega_D$ and $J$ (by setting $\omega_D^1=\omega_D^2$). Top panel: $N=2$. Middle panel: $N=5$. Bottom panel: $N=8$. Parameters used: $R=1$, $\tau_c=1$, $\Omega_1=15=\Omega_2$, $\Delta_1=1$, $\Delta_2=1.1$, $\varphi_1=\pi$.}
 	\label{Fig7}
 \end{figure}
 
\begin{figure}[h]
	\includegraphics[width=7.cm]{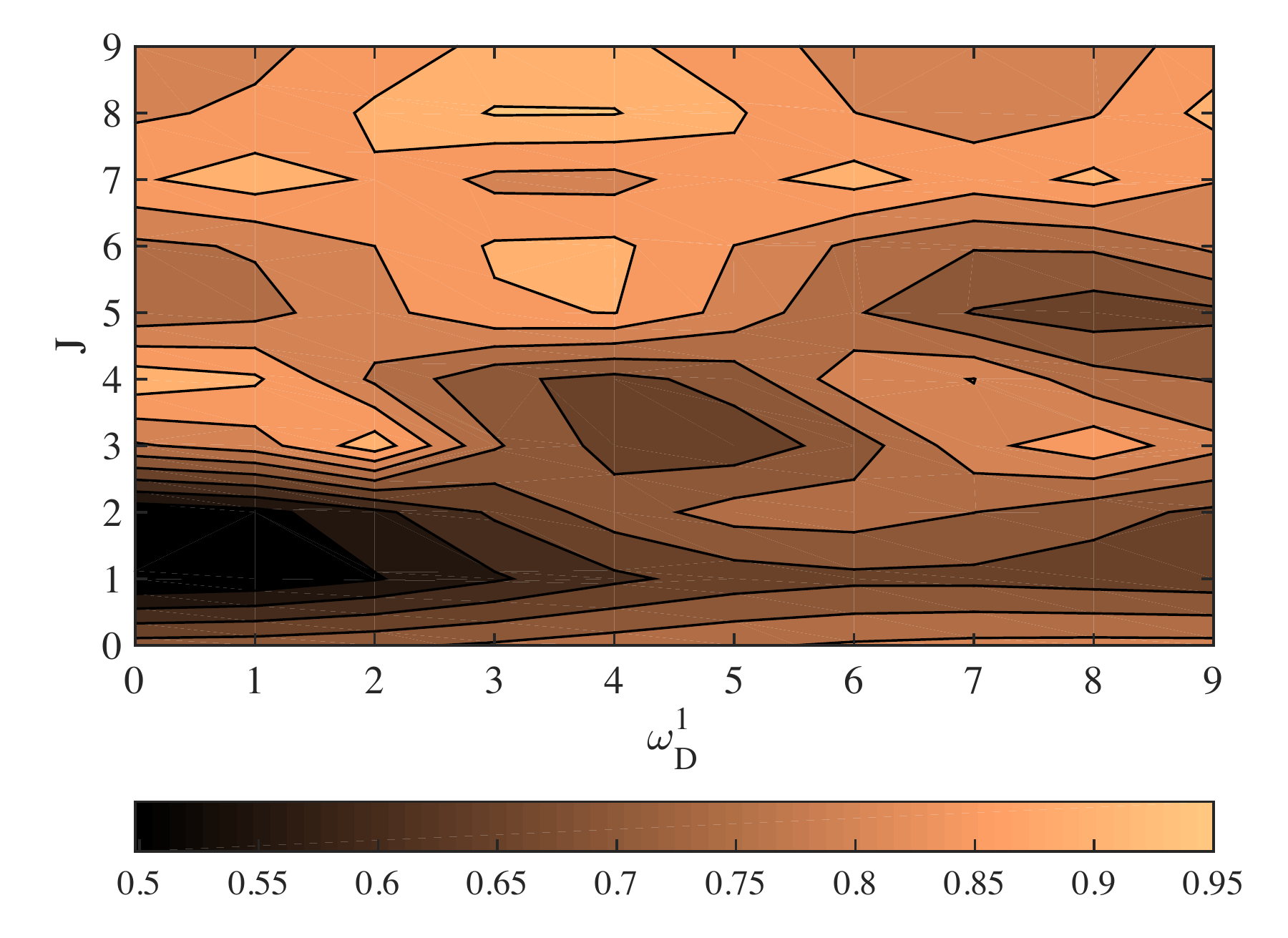}
 	\caption{(Color online) Concurrence as time evolves for an initially MES $|\Phi_-\rangle$ driven to a quasi-resonant case, for different values of  $\omega_D^1$ and $J$ (by setting $\omega_D^2=1$) and $N=2$. Parameters used: $R=1$, $\tau_c=1$, $\Omega_1=15=\Omega_2$, $\Delta_1=1$, $\Delta_2=1.1$, $\varphi_1=\pi$.}
 	\label{Fig7_2}
 \end{figure}

  In Figure \ref{Fig7} we show the Concurrence for a representative situation of an initial state $|\Phi_-\rangle$, where we show how it varies for different time sequences and different values of  $\omega_D$ and the transverse coupling $J$. We can see that, even though initially there are greater areas where Concurrence is null, as long as time evolves those regions become smaller. Quantum correlations seem to survive for big values of the transverse coupling among the particles of the bipartite system. There is an interesting pattern around $J\sim \omega_D$  due to ours choice of models parameters. In Figure 7, we have set $\omega_D^1=\omega_D^2=\omega_D$.  We can note that, in the case of equal frequencies $\omega_1=\omega_2=\omega$, $H_s|\Phi_-\rangle=\frac{-J+\omega}{\sqrt{2}}|\Phi_-\rangle$. As in that figure we have considered $\Omega_1=\Omega_2$ and $\Delta_1 \simeq \Delta_2$, it is easy to see that if $J \sim \omega_D$, we can assume the system is near another resonance. In order to exemplify this statement we have included in Figure \ref{Fig7_2}, the behaviour for the Concurrence as function of $J$ and $\omega_D ^1$, but setting  $\omega_D^2=1$ for $N=2$. Likewise, we can see that Concurrence has bigger values for larger values of $J$ for an equal time comparison. 
  
In this Section, we have shown that adding driving frequencies to two particles embedded in a structured environment can sometimes preserve quantum correlations. By comparing both initial bipartite states $|\Phi_+\rangle$ and $|\Phi_-\rangle$, we have seen that in both cases quantum correlations are better preserved for similar detuning frequencies rather than in a highly  biased detuned scenario. In the case of an initial MES $|\Phi_-\rangle$ state, due to the particular characteristics mentioned above, we have further seen that quantum correlations are better preserved when the initial  state is driven to a quasi resonant case. In addition, we have also shown that Concurrence become less robust as we get apart from the resonance condition. In those cases, the relation between $\Delta/\omega_D$ becomes important if the detuning frequencies are similar. Further, having an extra source of entanglement such as the transverse coupling considered herein, seems to enhance quantum correlations for this particular state.
 All in all, the quasi resonant case with initial bipartite state $|\Phi_-\rangle$ seems a particular scenario where driving can help to preserve quantum correlations.
 

\subsection{Initial maximally entangled states $|\Psi_{\pm}\rangle$}

\begin{figure}[h]
	\includegraphics[width=8cm]{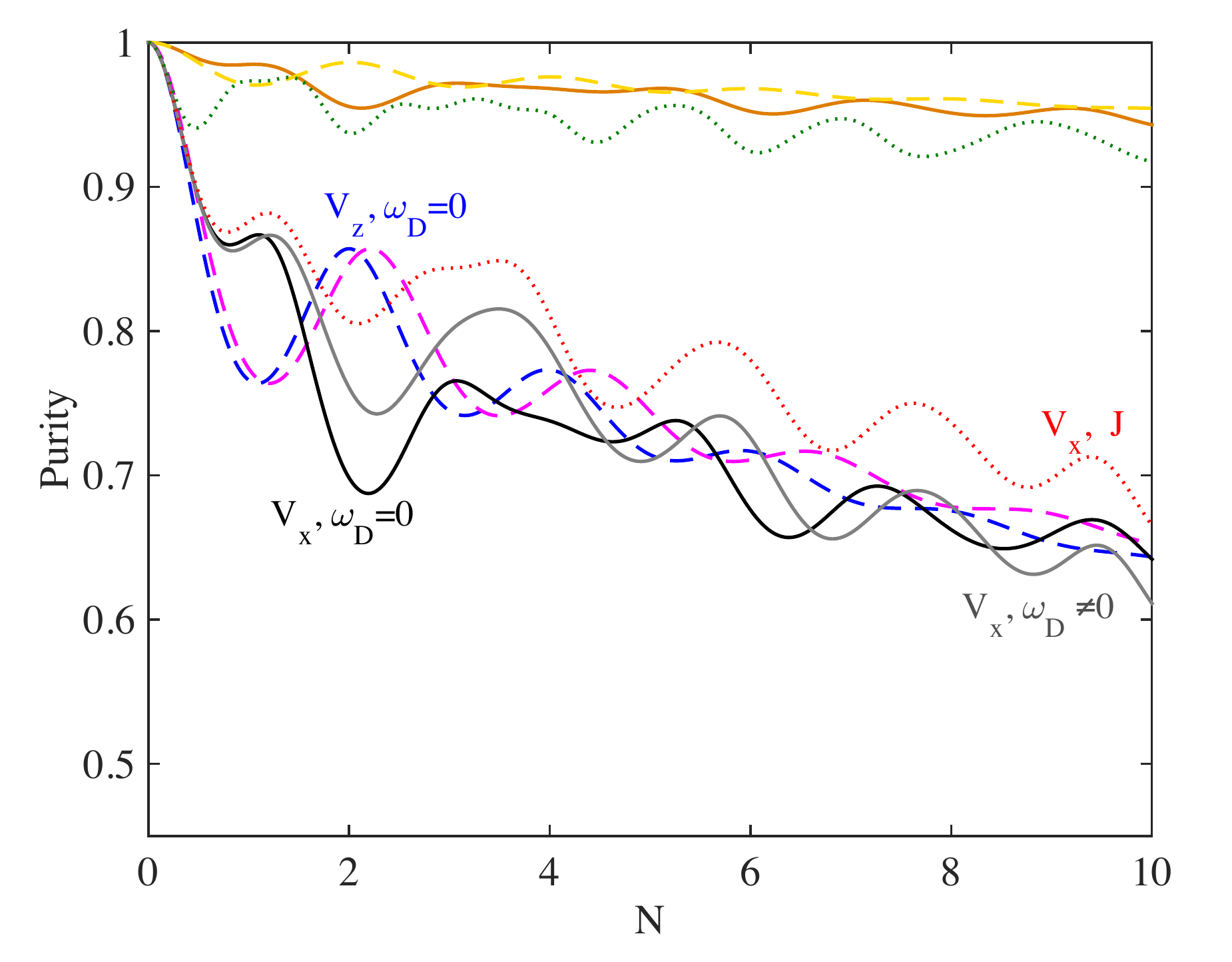}
 	\caption{(Color online)  Degradation of Purity for different coupling strength in both comparative models: solid lines represent the potential proportional to $\sigma_x$ while non-solid lines labeled with $V_z$ correspond to a purely dephasing model for weak $R=0.1$ and strong coupling $R=1$. For strong coupling $R=1$, we have further compared the loss of Purity for a static $\omega_D^{1,2}=0$ (black and blue lines) and non static situation $\omega_D^{1,2}\neq0$ (gray line for $\Delta_1=0.3$, $\Delta_2=0.3$, $\varphi_1=\pi$, $\omega_D^{1,2}=7$ while the pink line for $\Delta_1=0.4$, $\Delta_2=0.6$, $\varphi_1=\pi$, $\omega_D^{1,2}=3$) when comparing both interaction models. We have further added $J=1$ to the driven representative values for the $V_x$ model in the dotted red line  and $J=10$ for the green dotted line on top.
 	The orange solid and yellow dashed lines on top are curves for the undriven situation with $R=0.1$ for both models.
 	Parameters $R=1$, $\tau_c=1$,  $\Omega_1=\Omega_2=10$.}
 	\label{Fig11}
 \end{figure}

Let us now consider the case of two excitations initially present in the system. In this case, we can consider the initial state  limited to the subspace spanned by $\{|00\rangle,|11\rangle\}$, represented by $|\Psi_{\pm}\rangle$ of Eq. (\ref{Bell}) \cite{nota2}. As we have seen in the above section, there are some situations in which quantum correlations can be preserved. If this also happens for initial states $|\Psi_{\pm}\rangle$, we can search for further consequences as for example the acquisition of a geometric phase.
 The advantage of studying the dynamics of this maximally entangled state under the action of a strong environment is that it can be compared to some known results obtained for a purely dephasing model, 
say an interaction potential defined as $V_z=\sigma_z^1 \otimes  \mathbb{1}_2 + \mathbb{1}_1 \otimes  \sigma_z^2 $ (instead of the dipolar potential defined in Eq. (2). We can start by studying the loss of Purity as time evolves and compare among models. In Figure \ref{Fig11} we show some representative values for both interaction models: solid lines correspond to a dipolar coupling  defined in Eq. (2)  while dashed lines are for a purely dephasing model with $V_z$.
 We can see that for all couplings considered, say $R=0.1$ and $R=1$, solid lines decrease a greater quantity than the non-solid ones (for a dephasing model) when the parameters are similar. In the case of a strong coupling, we have also included the comparison among models when we consider the bipartite system to be driven, say $\omega_D^{1,2}\neq 0$.
 The magenta dashed line is the driven evolution for a dephasing model while the gray solid line is the equivalent under the dipolar coupling. The black solid line represents the static situation for the dipolar coupling while the blue dashed is the undriven evolution for the dephasing model. We can further notice that the effect of the driving is merely a shift in time in the case of the dephasing model. However, in the case of the potential considered in this manuscript, driving leads to a more interesting dynamics as has been already shown in the preceding section. In particular, Purity is enhanced by the inclusion of a transverse coupling (red dotted lines in Figure \ref{Fig11} ) among the particles (in addition to driving)  in the dipolar coupling. 
 In the following, we shall further focus on another interesting feature of bipartite system. It is well known that a MES state acquires a geometric phase of value $\pi$ (or $0$) and 0 for a separable state \cite{lucho}. For this reason, the term topological phase is generally used for the specific case of geometric phases acquired by maximally entangled states.
 The understanding of geometric phases (GPs) for entangled states is particular relevant due to potential applications in holonomic quantum computation with spin systems, which provide a plausible design of a solid-state quantum computer. The effect of the environment on a bipartite two-level system coupled to an external environment (bosonic or spin bath) was reported in \cite{bipartite, PRA83}. In particular, the GP correction for certain maximally entangled states  was shown to be null. That is, the phase is built as for unitary evolutions, as a stepwise behavior in steps of $\pi$. The study for a dephasing model was extended to two-qudit in \cite{annals}. In this framework, we shall inquire into the role of driving in preserving the geometric phase (or not) for the model presented in this manuscript and compare to the results of a dephasing model presented in \cite{bipartite}.
 
 A proper generalization of the geometric phase for
unitary evolution to a non unitary evolution is crucial for practical implementations of geometric
quantum computation. In \cite{Tong}, a quantum kinematic approach was proposed and the geometric phase 
(GP) for a mixed state
under non-unitary evolution has been defined as
\begin{eqnarray} 
\Phi & = &
{\rm arg}\{\sum_k \sqrt{ \varepsilon_k (0) \varepsilon_k (\tau_s)}
\langle\Psi_k(0)|\Psi_k(\tau_s)\rangle \nonumber \\ 
& & \times e^{-\int_0^{\tau_s} dt \langle\Psi_k|
\frac{\partial}{\partial t}| {\Psi_k}\rangle}\}, \label{fasegeo}
\end{eqnarray}
where $\varepsilon_k(t)$ are the eigenvalues and
 $|\Psi_k\rangle$ the eigenstates of the reduced density matrix
$\rho_{\rm r}$ (obtained after tracing over the reservoir degrees
 of freedom). In the last definition, $\tau_s$ denotes a time
after the total system completes a cyclic evolution when it is isolated from the environment \cite{nota2}. We can compute the GP acquired by the bipartite system since we have numerically solved the dynamics the bipartite system. When the system is open, the geometric phase that would have been obtained if the system had been closed, $\Phi_u$, is modified. This means, in a general case, the phase is $\Phi_g = \Phi_u + \delta \Phi$, where $\delta \Phi$ is the correction to the unitary phase, induced by the presence of the environment.
 
\begin{figure}[h]
	\includegraphics[width=8cm]{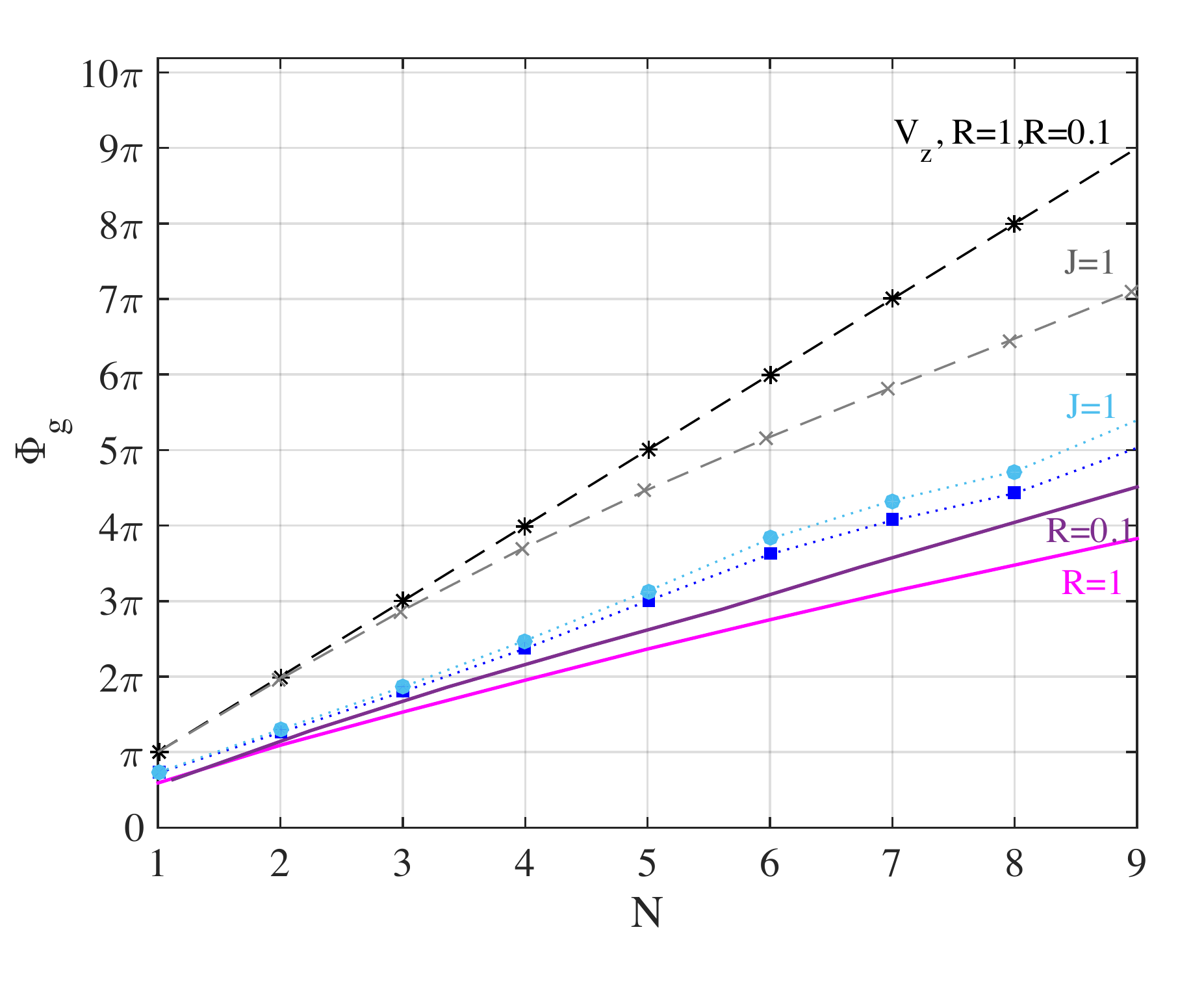}
 	\caption{(Color online) Geometric phase acquired by the bipartite state (initially MES) while evolving under different environment parameters. The black asterisk-dotted line is the topological phase acquired by the MES state for a purely dephasing model, while considering different environments strengths or even driving: it is always $\pi$. The gray dashed line with crosses is for a driven representative situation in the dephasing model $\omega_D^{1}=2$ and $\omega_D^{2}=3$ when there is  interaction among particles $J=1$.  All other lines are for the geometric phase of the MES state for the interaction dipolar potential $V$ (proportional to $\sigma_x$). Solid lines (purple line $R=0.1$ and pink line $R=1$) represent no-driving situations while dotted lines represent different driving parameters for a strong coupling regime:  blue line with squares represents $\omega_D^{1,2}=7$ while the light blue line with circles is for $\omega_D^{1,2}=7$ and $J=1$. Parameters used: $\tau_c=1$, $\Omega_1=10=\Omega_2$, $\Delta_1=0.3$, $\Delta_2=0.3$, $\varphi_1=\pi$.}
 	\label{Fig12}
 \end{figure}
 
 In Figure \ref{Fig12} we show the geometric phase acquired by the bipartite state (initially MES) while evolving under different environment parameters. The black asterisk-dashed line is the topological phase acquired by the MES state for a purely dephasing model $[H_s,H_I]=0$, while considering different environments or even driving: it is always $\pi$. All other lines are for the geometric phase of the MES state for the dipolar interaction potential $V$ proportional to $\sigma_x$ defined in Eq. (2). Both solid lines represent the geometric phase obtained in the static case, say $\omega_D^{1,2}=0$ for $R=1$ and $R=0.1$.  Dotted lines with circles and squares represent the driven situations for $R=1$ and $J=1$. 
 In an unitary evolution, this geometric phase has a topological nature, and it has been proved to be robust in open quantum evolution under dephasing models \cite{bipartite, annals}.  In this latter case, the geometric phase acquired by the bipartite system is $\pi$ when the initial state is a maximally entangled. However, when the interaction hamiltonian is dipolar, the model is not longer a dephasing one and  the geometric phase looses this topological nature. Hence, it becomes sensitive to external noise.
Another important feature we can note is the fact that the GP for the undriven case under a strong evolution is approximately $\pi/2$ and then it evolves acquiring a geometric phase proportional to that value in each cycle (solid  lines). In \cite{PRA83} some of us have computed the geometric phase acquired by a two level system under the presence of a composite environment, formed by an external bath and another two level particle. It has been reported that when the initial entanglement among the two spin 1/2 particles was maximal then the geometric phase acquired by one spin 1/2 particle was $\pi/2$, differently to the geometric phase acquired by the bipartite system evolving under the presence of the external bath (dephasing model). 
 \begin{figure}[h]
	\includegraphics[width=8cm]{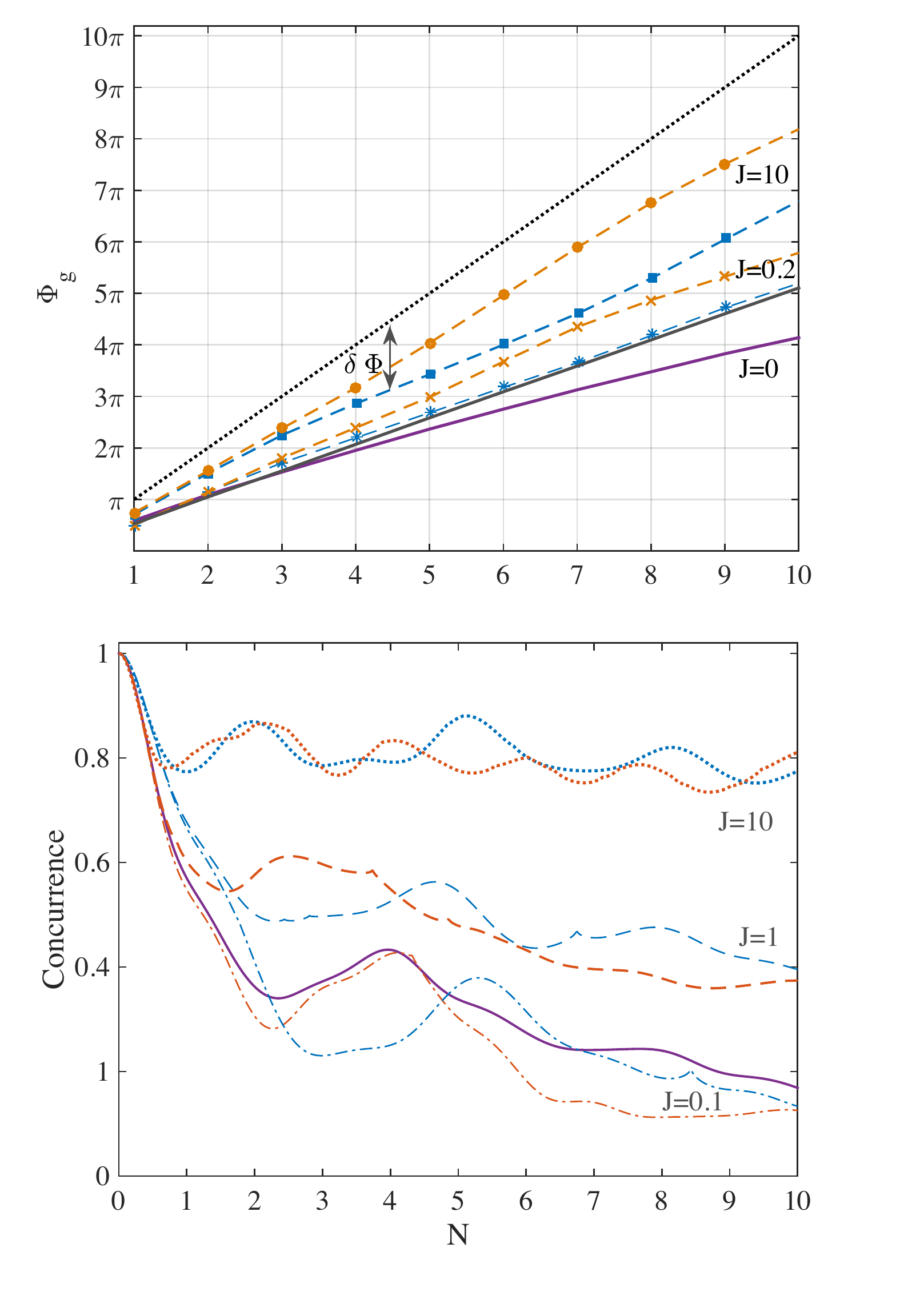}
 	\caption{(Color online) Top: Geometric phase acquired by the bipartite state (MES) while evolving under different environment parameters. The black dotted line is for topological phase acquired by the bipartite. The purple solid line represents the phase acquired under a dipolar evolution with $J=0$.  The dotted-dashed lines with crosses and asterisks are the corresponding phases acquired when $J=0.2$. Finally, we include the phase acquired with $J=10$ for the corresponding evolution in the dashed lines with circles and squares. Blue lines represent equal $\Delta$ values, while orange lines are for opposite $\Delta$ values as explained in bottom panel. The dark grey solid line correspond to $J=50$. Bottom: Concurrence evolving in time for the driven evolutions for different values of $J$: dot-dashed correspond to $J=0.1$, dashed lines to $J=1$ and dotted lines to $J=10$.
 	 The blue and orange lines are for different representative driven situations: orange lines for $\Delta_1=3$, $\Delta_2=3$, $\varphi_1=\pi$, $\omega_D^{1}=2$ and $\omega_D^{2}=3$ and blue lines for $\Delta_2=3=\Delta_1$ and $\varphi_1=0$, $\omega_D^{1}=2$ and $\omega_D^{2}=3$. The purple line is for $J=0$ for the undriven dipolar coupling.
 	Parameters: $\tau_s=1$, $\Omega_1=10=\Omega_2$.}
 	\label{Fig13}
 \end{figure}
The effect of a dipolar coupling on the geometric phase can be compared as to the one induced by a composite environment  \cite{PRA83}. Further, we can note that the geometric phase acquired for a driven bipartite system is less corrected  than for the equivalent undriven situation ($\delta \Phi|_{\omega_D} < \delta \Phi|_{\omega_D=0}$): the driving done in a strong regime $R=1$ (dotted lines with squares) renders similar values of the geometric phase acquired in an evolution under a weaker environment $R=0.1$ (undriven evolution purple solid line).
This result agrees with a previous one, where it has been reported that 
some values of  driving ``preserves" the geometric phase \cite{PRAdriven}.
Finally, we can also analyze the role of the transverse coupling in the geometric phase acquired. In the case of the dephasing model, we can see that adding a transverse coupling destroys the robust condition for long time evolution (gray dashed  line with crosses in Figure \ref{Fig12}). Initially, we obtain a similar value to the topological one $\phi_g(N=1)=\pi$ but at later times, the loss of information towards the environment corrects significantly the topological value obtained when isolated. However, for a dipolar environmental coupling (dotted line with dots), we can note that its effect is different, as it leads to  a smaller correction to $\Phi_u$ even yet when compared to a non-driven evolution under a weaker coupling regime $R=0.1$.

In Figure \ref{Fig13} we can see the geometric phase acquired for the MES under a dipolar coupling for a driven evolution and different values of the dimensionless parameter $J$.
Therein, we can see a representative undriven situation in a purple solid line for $J=0$. As it can be seen, when compared to the robust topological GP (black dotted line), it is evident the strong effect of the environment during the evolution (recall that in the case of the dotted black line $\delta \Phi=0$, since $\Phi_g=\Phi_u$. The greater the deviation from this line, the bigger $\delta \Phi$ for each evolution is). As we increase the value of the transverse coupling we can notice that the correction to the unitary geometric phase decreases even though the system evolves under a strong coupling. All dashed lines represent driven situations for different values of $J$. The two different color are included to identify different detuning situations: orange for $\Delta_1=-\Delta_2$ and $\Delta_1=\Delta_2$, such that $\omega_D/\Delta \sim {\cal O}(1)$. This driving has been shown to suppress non-markovianity effects \cite{Poggi,PRAdriven}. The contribution of the transverse coupling towards a preservation of the GP is considerable. This situation can be explained by the fact that this parameter contributes to preserving the quantum correlations as can be seen in the bottom panel of  Figure \ref{Fig13}. As the Purity of the system remains approximately constant, the geometric phase acquired in each cycle is approximately constant as well. We can further notice a gray solid line that represents $J>>1$. In that case, Purity is closer to one along the evolution. Initially ($N=1$), geometric phase is $\Phi_g=\pi/2$ as reported in \cite{PRA83} and continues constant (almost a straight line like the dotted one but with a different rate) during the evolution even though the bipartite is in a non-markovian environment in a strong regime. 
\section{Conclusions}
\label{conclusions}

In this manuscript we have studied the quantum correlations of a driven bipartite state embedded in a common structured environment for different regimes. We have focused the analysis on maximally entangled states of X-class bipartite state. We have implemented a hierarchy equations numerical method since it can describe the dynamics of the bipartite system with a nonperturbative and non-markovian system-bath interaction at finite temperature, even under strong time-dependent perturbations. This formalism is valuable because it can be used to study not only strong-bath coupling, but also quantum coherence or quantum entanglement. The information concerning the system-bath coherence is stored in the hierarchical elements, which allows us to simulate the quantum entangled dynamics between the system and the environment.

In the case of only one excitation present in the system and in the environment, we have studied the degradation of Purity of the bipartite system for different coupling regimes.  We considered two different particles  (but similar frequencies) and studied the quantum correlations (measured by the Concurrence) if the particles were driven out and on resonance, so as to established in which case quantum correlations were longer preserved. We showed that an initial MES of the form $|\Phi_-\rangle$  driven to a quasi-resonant situation preserves longer quantum correlations and that this effect could be enhanced by the presence of a transverse coupling among the particles of the bipartite system.

As for the case of zero and two excitations in the system and environment, we further dug into the effect of driving and transverse coupling on the geometric phase acquired during the evolution. We have compared the results obtained in the model with those reported in the literature for a simpler model of dephasing. We have shown that the robustness condition of the geometric phase is lost when the interaction potential is dipolar. In the case of the dephasing model, we can see that adding a transverse coupling destroys the robust condition for long time evolutions.  However, for a dipolar coupling with the environment, we have noted that its effect is different.  The dipolar coupling does not preserve the robustness of the geometric phase even though the initial state is a maximally entangled one. The transverse coupling among the particles composing the bipartite was shown to help extend quantum correlations in time leading to  a smaller correction of the unitary geometric phase (even yet when compared to a non-driven evolution under a weaker coupling). A combination of driving and transverse coupling among the particles can help suppress the non markovian effects. 
The dipolar coupling does not preserve the robustness of the geometric phase even though the initial state is a maximally entangled one. Instead, the environmental induced dynamics modifies coherences and populations of the reduced density matrix leading to a great variation of the GP. However, if we introduce a driving into the bipartite system and a transverse coupling among the two-level particles, we can obtain a constant geometric phase, which is not the topological geometric phase. It rather corresponds to the geometric phase acquired by one particle  initially composing a  maximal entangled state coupled to a spin boson (dephasing model) when the degrees of freedom of the environment and one of the particles are traced out. 
 
 The possibility of exploiting the environment as a resource for control has opened a new door in the manipulation of open quantum systems. The generation and stabilization of entanglement is one of the main challenges for quantum information applications.  The  model presented in this manuscript can be used to simulate experimental situations such as hybrid quantum classical systems feasible with current technologies.  
 It is important to note that if the noise effects induced in the system are of considerable magnitude, the coherence terms of the quantum system are rapidly destroyed and the GP literally disappears. It has been argued that the observation of GPs should be done for times long enough to obey the adiabatic approximation but short enough to prevent decoherence from deleting all phase information. As the geometric phase accumulates over time, its correction becomes relevant on a relatively short timescale, while the system still preserves purity. All the above considerations lead to a scenario where the geometric phase can still be found and it can help us infer features of the quantum system that otherwise might be hidden to us.\\
\section*{Acknowledgements}
We acknowledge UBA, CONICET and ANPCyT--Argentina.   
The authors wish to express their gratitude to the TUPAC cluster, where the calculations of this paper have been carried out. We thank F. Lombardo his discussions and comments. P.I.V acknowledges ICTP-Trieste Associate Program.\\

\appendix
\section{Dynamical evolution for an initial bipartite state $|\Phi_+\rangle$ }
\label{appendixa}

The dynamics of the bipartite will depend on three ingredients: driving, coupling and dissipation.  In the following, we shall define all timescales involved in the evolution of the bipartite state. The bath correlation time is defined as $t_c= \lambda^{-1}$. As Eq. (9) deals with re-scaled quantities, $\tau_c=1$ for this manuscript. The relaxation timescale is then determined  by $\tau_r=\gamma_0^{-1}$.
As for the bipartite system itself, in the case of one excitation present in the system,  there is a characteristic timescale defined as $\tau_s=2\pi/(\omega_2-\omega_1)$, that is the time at which the closed bipartite system will get back to its initial state. For zero or two excitations present in the system, the characteristic time is set as $\tau_s=2\pi/(\omega_1+\omega_2)$.
Finally, if it were a dephasing model, we could define the characteristic decoherence timescale $\tau_D$, as the estimated time at which the coherences are dynamically suppressed by the presence of the environment. As the model defined herein is not a dephasing one, we can get an insight into the Purity of the bipartite state so as to have an estimation of the degradation of Purity suffered by the interaction with an environment.

Herein, we shall consider a structured environment and have no limitations on the coupling. The factor $R$ is defined as the rate between the coupling strength between the system and the bath $\bar{\gamma}_0$ and $\lambda$,  defined as the
broadening of the spectral peak of the environment. Particularly, we are interested on a strong coupling $R=1$. By assuming $R=1$, we are implying $\bar{\gamma}_0 \sim \lambda$, which means that $\tau_r \sim \tau_c$.
In such a case, non-Markovian dynamics induced by the reservoir memory  (describing the feedback of information and/or energy from the reservoir into the system) becomes important. 

\begin{figure}[h]
	\includegraphics[width=8cm]{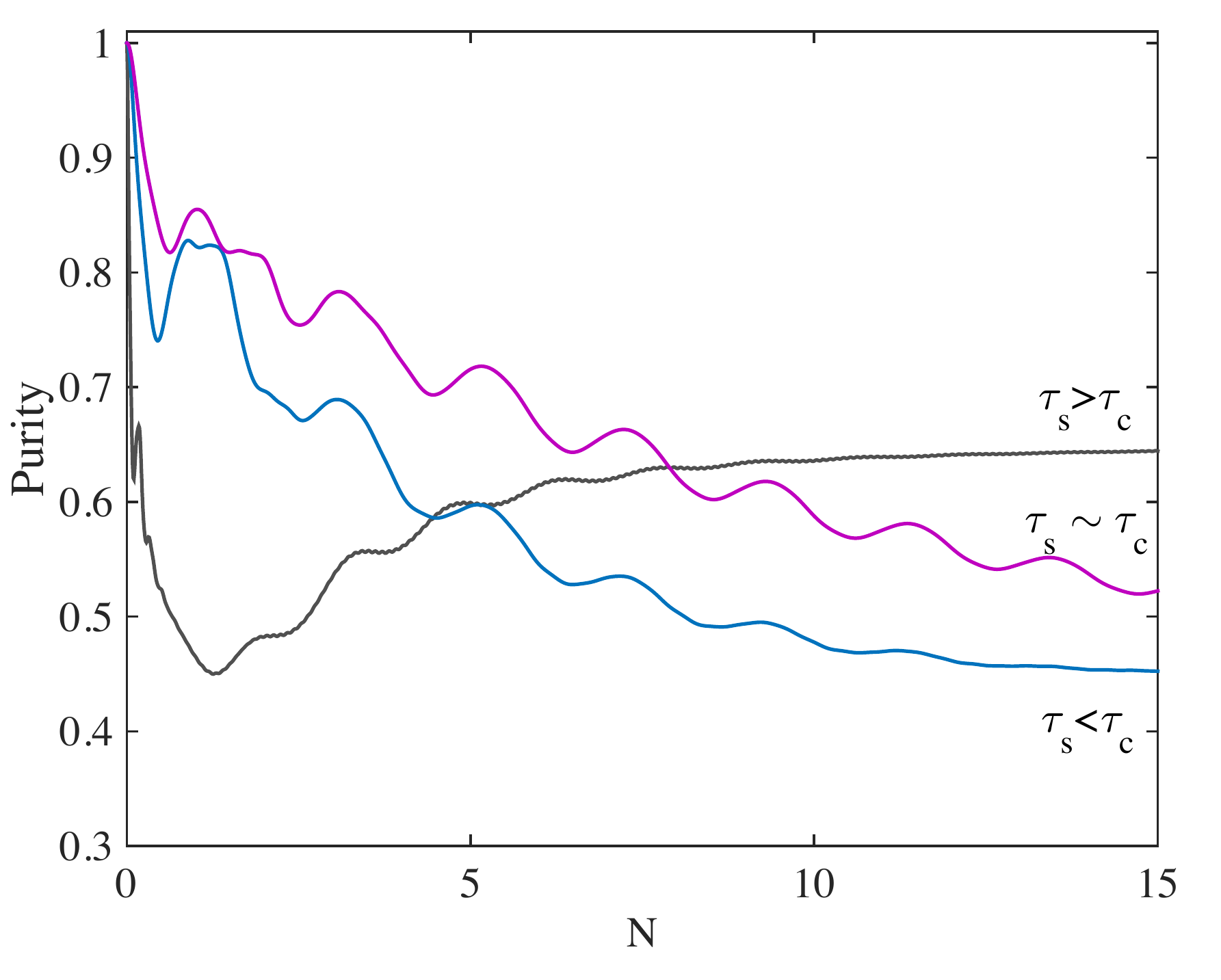}
 	\caption{(Color online)  Loss of Purity for three different bipartite systems for $\tau_s < \tau_c$, $\tau_s \sim \tau_c$ and $\tau_s > \tau_c$. The magenta line is for $\Omega_2=16$, $\Omega_1=10$. The blue line is for $\Omega_2=20$, $\Omega_1=10$ and the gray one for $\Omega_2=11$, $\Omega_1=10$. In all cases the environment exhibits a strong coupling $R=1$, $\tau_c=1$, $\Delta_1=0$, $\Delta_2=0$.}
 	\label{apendice1}
 \end{figure}

In this section, we shall go over on some issues 
about the dynamical evolution of the bipartite in one-excitation subspace for an initial bipartite state $|\Phi_+\rangle$, having defined a strong coupling with the environment.
In Figure \ref{apendice1} we show three different situations: $\tau_s < \tau_c$, $\tau_s \sim \tau_c$ and $\tau_s > \tau_c$ ($\tau_c=1$). Therein, we can note that if we want to study the effect of driving during the dynamical evolution, it is important to choose the model's parameters that allow such investigation. In the case of $\tau_s > \tau_c$, the system is near resonance since frequencies are extremely similar. On the contrary, for $\tau_s < \tau_c$ frequencies are very different. 
The qualitatively different behaviours at long times must be understood by recalling the fact that $N=\tau/\tau_s$ ($\tau$ the dimensionless time defined in Eq. (\ref{hierarchy})). As  $\tau_s$ varies, time elapsed $\tau$ for N cycles is different for the different set of parameters. This means that, for example at $N=15$, the bipartite evolution can be found at dissimilar stages of the dynamical evolution for the different representative lines.

\begin{figure}[h]
	\includegraphics[width=8cm]{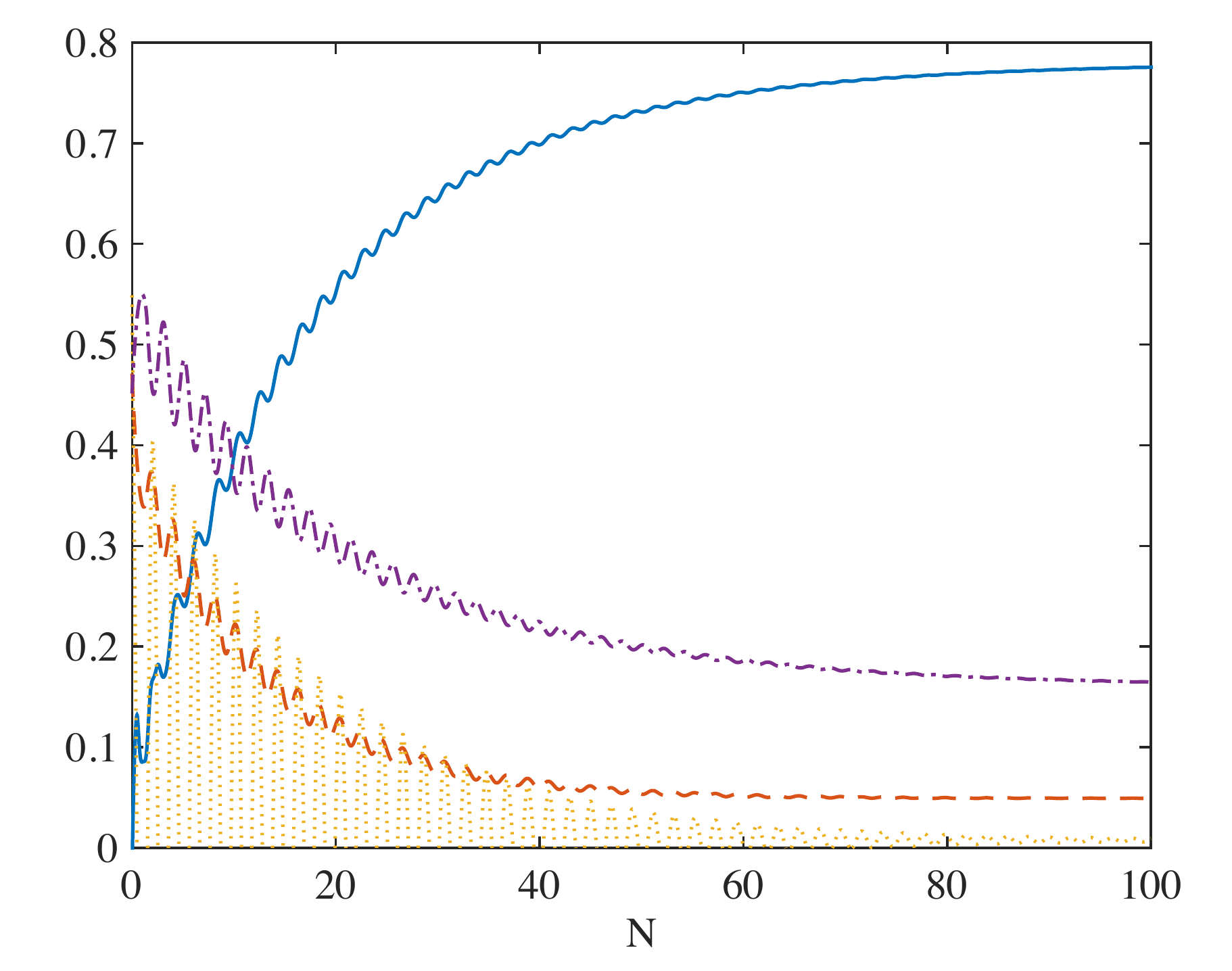}
	\caption{(Color online) Dynamical evolution for some of the elements of the reduced density matrix defined in Eq. (\ref{rho2})  for $\tau_s \sim \tau_c$: dot-dashed purple line represents $\rho_{22}(t)$, red dashed line $\rho_{11}(t)$ while the blue solid line for $\rho_{00}(t)=1-\rho_{11}(t)-\rho_{22}(t)$. The orange dotted line represents $Re[\rho_{21}(t)]$.
	We can see that for these parameter's values the system reaches its asymptotic values for approximate $N=100$. 
 	Parameters: $R=1$, $\tau_c=1$, $\Omega_1=10$, $\Omega_2=16$. No driving is considered.}
 	\label{apendice2}
 \end{figure}
\begin{figure}[h]
	\includegraphics[width=8cm]{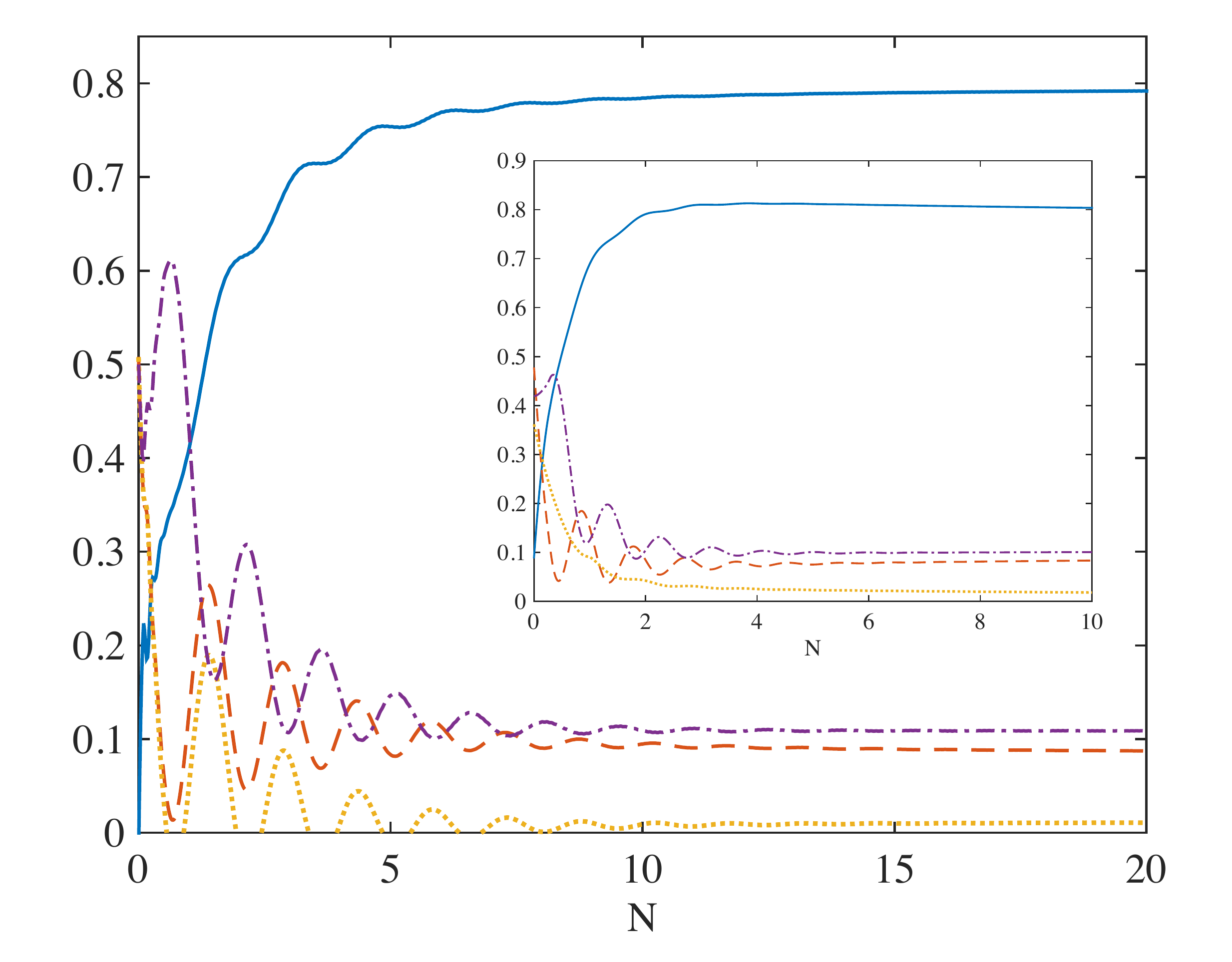}
	\caption{(Color online) Dynamical evolution for some of the elements of the reduced density matrix defined in Eq. (\ref{rho2})  for $\tau_s > \tau_c$: dot-datshed purple line represents $\rho_{22}(t)$, red dashed line $\rho_{11}(t)$ while the blue solid line for $\rho_{00}(t)=1-\rho_{11}(t)-\rho_{22}(t)$. The orange dotted line represents the $Re[\rho_{21}(t)]$.
	We can see that for these parameter's values the system reaches its asymptotic values for approximate $N=10$. 
 	Parameters: $R=1$, $\tau_c=1$, $\Omega_1=10$, $\Omega_2=11$. No driving is considered. Inset: Dynamical evolution for some of the elements of the reduced density matrix for more similar frequencies $\Omega_1=10$, $\Omega_2=10.5$. The decay of the coherences is faster. }
 	\label{apendice3}
 \end{figure}

In order to explain the qualitatively different behaviors for $\tau_s \sim \tau_c$ and $\tau_s > \tau_c$, we have further included Figure \ref{apendice2} and Figure \ref{apendice3}.

 In Figure \ref{apendice2} we show the dynamical evolution of the representative matrix elements Eq.(\ref{rho2}) for $\tau_s \sim \tau_c$. The
 dot-dashed purple line represents $\rho_{22}(t)$, the red dashed line $\rho_{11}(t)$ and the blue solid line $\rho_{00}(t)=1-\rho_{11}(t)-\rho_{22}(t)$. The orange dotted line represents the real part of the coherences $\rho_{21}(t)$.
 We can see therein that the asymptotic state is achieved for approximately $N=100$ and the coherences oscillate for several values of $N$. For the temporal timescales studied in Section II.A, we can note the typical oscillations of the Non-Markovian evolution ruled by a strong interaction. Oppositely, in Figure \ref{apendice3} we show the dynamical evolution of the representative matrix elements Eq.(\ref{rho2}) for a quasi-resonant condition, $\tau_s > \tau_c$. 
 We can note that the asymptotic state is reached at a fewer number of $N\sim 10$ and coherences oscillate for only a few cycles. In the inset, we can see another set of parameters, where frequencies are even closer and $\tau_s>>\tau_c$. Therein, we can see the asymptotic state is reached for $N\simeq 2$ and coherences decay rapidly. In time units, we can see that decoherence process is accelerated near the resonance condition.
 These numerical results can be understood by the help of some analytical estimations obtained under the rotating wave approximation when considering the resonance case, say $\omega_1=\omega_2$. In that case, authors showed that coherences decay with a rate set by the environmental spectral broad $\lambda$ \cite{PRL100}. 
 
 In our manuscript, we particularly consider the situation where particles have similar (but not equal) frequencies. Therefore, the model allows to drive the particles in and out resonance and see if there is a particular situation where quantum correlations are enhanced. In order to compare equal-timescale evolutions, we ask for $\tau_s \sim \tau_c$. Adding driving will necessary add another timescale to the already complex evolution. Dynamical evolution is different for an initial bipartite state $|\Phi_-\rangle$ near resonance since $|\Phi_-\rangle$ is an eigenstate of both $H_s$ and $H_I$ when $\omega_1=\omega_2$ \cite{PRA85}.

{}


\begin{thebibliography}{plain}
\bibitem{nota} In the resonant case $\omega_1=\omega_2=\omega$, we have defined $N=\tau/\tau_s$, being $\tau_s$ an arbitrary definition $\tau_s =2\pi/\omega$.
\end{thebibliography}
\end{document}